\documentclass[conference]{ieeeconf}
\IEEEoverridecommandlockouts
\usepackage{amsmath, amssymb, mathrsfs}
\usepackage{mathtools}
\usepackage{graphicx}
\usepackage{xcolor}
\usepackage{multirow}
\usepackage{enumerate, paralist}
\usepackage{booktabs}
\usepackage{afterpage}
\usepackage{cite}

\usepackage{tikz}
\usetikzlibrary{calc,shapes,arrows,positioning}
\usetikzlibrary{patterns, decorations.pathreplacing, decorations.pathmorphing, decorations.markings}

\usepackage{algorithm}
\usepackage[noend]{algorithmic}

\usepackage{amsthm}
\usepackage{refcount}

\newtheorem{proposition}{Proposition}
\newtheorem{lemma}[proposition]{Lemma}
\newtheorem{corollary}{Corollary}[proposition]

\theoremstyle{definition}
\newtheorem{definition}[proposition]{Definition}

\theoremstyle{remark}
\newtheorem{remark}[proposition]{Remark}

\newcommand{\real}{\mathbb{R}}

\newcommand{\integer}{\mathbb{Z}}
\renewcommand{\natural}{\mathbb{N}}
\newcommand{\norm}[1]{\left\lVert{#1}\right\rVert}
\DeclareMathOperator*{\argmin}{argmin}
\DeclareMathOperator{\rank}{rank}

\newcommand{\Sys}{{\mathcal{S}}}
\newcommand{\liftvec}[1]{\mathbf{#1}}
\newcommand{\liftmat}[1]{\mathfrak{#1}}
\newcommand{\LiftSys}{{\mathcal{S}_\mathsf{L}}}
\newcommand{\lag}{\mathbf{l}}
\newcommand{\order}{\mathbf{n}}
\newcommand{\Idx}{\theta}
\newcommand{\ProperIdx}{\Theta}
\newcommand{\EstProperIdx}{\widehat\Theta}
\newcommand{\prev}[1]{\mathrm{prev}(#1)}

\newcommand\oprocendsymbol{\hbox{$\square$}}
\newcommand\oprocend{\relax\ifmmode\else\unskip\hfill\fi\oprocendsymbol}

\newenvironment{pfof}[1]{\vspace{1ex}\noindent{\itshape Proof of
    #1:}\hspace{0.5em}} {\hfill\oprocend\vspace{1ex}}

\begin{document}

\title{Data-Driven Model Predictive Control for Linear Time-Periodic Systems}

\author{Ruiqi Li, John W. Simpson-Porco, and Stephen L. Smith
\thanks{This research is supported in part by the Natural Sciences and Engineering Research Council of Canada (NSERC).}%
\thanks{Ruiqi Li and Stephen L. Smith are with the Electrical and Computer Engineering at the University of Waterloo, Waterloo, ON, Canada
{\tt\small \{r298li,stephen.smith\}@uwaterloo.ca}}%
\thanks{John W. Simpson-Porco is with the Department of Electrical and Computer Engineering at the University of Toronto, Toronto, ON, Canada
{\tt\small jwsimpson@ece.utoronto.ca}}
}

\maketitle
\thispagestyle{empty}
\pagestyle{empty}

\begin{abstract}
We consider the problem of data-driven predictive control for an unknown discrete-time linear time-periodic (LTP) system of known period. Our proposed strategy generalizes both Data-enabled Predictive Control (DeePC) and Subspace Predictive Control (SPC), which are established data-driven control techniques for linear time-invariant (LTI) systems. The approach is supported by an extensive theoretical development of behavioral systems theory for LTP systems, culminating in a generalization of Willems' fundamental lemma. Our algorithm produces results identical to standard Model Predictive Control (MPC) for deterministic LTP systems. Robustness of the algorithm to noisy data is illustrated via simulation of a regularized version of the algorithm applied to a stochastic multi-input multi-output LTP system.
\end{abstract}

\section{Introduction}

Control design methods can broadly be classified into \emph{model-based} methods and \emph{data-driven} methods.
Model-based design methods rely on an accurate parametric representation of the system, which may come from first-principles modeling or from system-identification.
Data-driven control, on the other hand, produces a control strategy directly from recorded historical data.
As modern systems of interest become increasingly complex and difficult to identify, 
data-driven control techniques become increasingly preferable, and have attracted significant research interest in recent years.
A comprehensive survey of early data-driven methods can be found in \cite{INTRO:Hou2013}.

\emph{Model Predictive Control (MPC)} is a particular model-based design method which has been widely used in industrial applications,
such as autonomous driving \cite{INTRO:APP:Brown2017} and mobile robots \cite{INTRO:APP:Bullo2011}.
MPC is applicable to time-varying systems, and can incorporate input and output constraints, which typically model actuator saturation and safety constraints, respectively.
Despite these benefits, MPC requires a parametric system model, and the modeling process (e.g. system identification) can be expensive \cite{INTRO:Ogunnaike1996}.

Originating from the work by J. C. Willems, \emph{behavioral systems theory} provides an alternative to the now-standard state-space framework \cite{BEHAV:Willems2007_survey}. 
In the behavioral approach, a system is characterized as a set of possible input-output trajectories (the \emph{behavior}). Of particular note, the behavior of a finite-dimensional discrete-time linear time-invariant (LTI) system over a finite time interval can be expressed using collected historical data, a result now known as the \emph{Fundamental Lemma} \cite{BEHAV:Willems2005}.
See \cite{BEHAV:Markovsky2021} for an introduction of the behavioral approach with relevant algorithms, and \cite{BEHAV:Willems2007} for perspectives of developing the theory.
Currently, many results in the behavioral framework are restricted to LTI systems.

By leveraging the fundamental lemma, system outputs can be predicted without a parametric system model. From this observation, \emph{data-driven MPC (DDMPC)} methods have been developed wherein the need for a parametric system model is eliminated \cite{INTRO:DDMPC:Berberich2020a, INTRO:DDMPC:Berberich2020b, INTRO:DDMPC:Berberich2021, Coulson2019, INTRO:Coulson2019b, INTRO:Coulson2021, INTRO:Huang2021a, INTRO:Huang2021b}.
A particular DDMPC algorithm named \textbf{D}ata-\textbf{e}nabl\textbf{e}d \textbf{P}redictive \textbf{C}ontrol (DeePC) \cite{Coulson2019, INTRO:Coulson2019b, INTRO:Coulson2021, INTRO:Huang2021a, INTRO:Huang2021b} has been successfully applied to control problems in power systems \cite{INTRO:APP_DeePC:Huang2019a, INTRO:APP_DeePC:Huang2019b}, motor drives \cite{INTRO:APP_DeePC:Carlet2020} and quad-copters \cite{INTRO:APP_DeePC:Elokda2021}.
While the methods \cite{INTRO:DDMPC:Berberich2020a, INTRO:DDMPC:Berberich2020b, INTRO:DDMPC:Berberich2021, Coulson2019, INTRO:Coulson2019b, INTRO:Coulson2021, INTRO:Huang2021a, INTRO:Huang2021b} focus on LTI systems, 
some extensions have been developed for linear parameter-varying systems \cite{INTRO:EXTENSION:Verhoek2021} and for specific types of nonlinear systems \cite{INTRO:EXTENSION:Luo2018}. 
In the spirit of these extensions, the extension to linear time-varying (LTV) systems is also of interest, and our focus here will be on developing analogous theory and control techniques for \emph{linear time-periodic (LTP)} systems --- a particular class of LTV systems. LTP systems can arise from linearization of nonlinear systems around periodic trajectories, such as in models of helicopters \cite{INTRO:APP_LTP:Verdult2004} and wind turbines \cite{INTRO:APP_LTP:Allen2011} etc. 

\emph{Contributions:} This paper develops behavioral systems theory and associated DDMPC results for LTP systems of known periods. Our key insight is that an established lifting technique (see \cite{INTRO:LIFT:Bittanti2000, LIFT:Jemaa2003} and Section~\ref{SEC:Lift}) which transforms an LTP system into an LTI system can be leveraged to extend the behavioral theory of LTI systems to LTP systems. Based on this, in Section~\ref{SEC:Beh} we develop the behavioral theory for LTP systems \textemdash{} generalizing notions such as order and lag \textemdash{} and culminating in a natural extension of the fundamental lemma \cite{BEHAV:Willems2005}. Leveraging this theory, in Section \ref{Sec:DDMPC} we put forward a DDMPC algorithm for LTP systems, generalizing the established DeePC and Subspace Predictive Control (SPC) \cite{INTRO:APP_DeePC:Huang2019a, INTRO:APP_DeePC:Huang2019b, INTRO:APP_DeePC:Carlet2020} methods for LTI systems. We provide a performance guarantee that for deterministic LTP systems, our algorithm gives the same control policy as obtained from MPC. Finally, we illustrate the effectiveness of our approach via a simulation study in Section~\ref{SEC:Simu}.

\emph{Notation:} 
    Let $[M_1; \ldots; M_k] := [M_1^\top, \ldots, M_k^\top]^\top$ denote the column concatenation of matrices $M_1, \ldots, M_k$.
Given a $\real^q$-valued discrete-time signal $z$ with an integer index, for integers $t_1,t_2$ with $t_1 \leq t_2$, let $z_{[t_1, t_2]}$ (resp. $z_{[t_1, \infty)}$) denote either the sequence $\{z_{t}\}_{t=t_1}^{t_2}$ (resp. $\{z_{t}\}_{t=t_1}^\infty$) or the concatenated vector $[z_{t_1}; \ldots; z_{t_2}] \in \real^{q(t_2-t_1+1)}$ (resp. the semi-infinite vector $[z_{t_1}; z_{t_1+1}; \ldots]$). 
Similarly, for integers $t_1 < t_2$, let $z_{[t_1, t_2)} := z_{[t_1, t_2-1]}$.
Let $M^\dagger$ denote the pseudo inverse of a matrix $M$.

All proofs in this paper are postponed to Appendix.
\setlength{\abovedisplayskip}{.3em}
\setlength{\belowdisplayskip}{.3em}

\section{Linear Time-Periodic Systems and the Lifting Technique}
\label{SEC:Lift}

In this section we review some classical notions for linear time-periodic systems.
Consider a discrete-time linear time-varying (LTV) system
\begin{equation}\label{Eq:LTV} \Sys:\bigg\{ \begin{aligned}
    x_{t+1} =&\; A_t x_t + B_t u_t\\
    y_t =&\; C_t x_t + D_t u_t
\end{aligned}
\end{equation}
with initial time $t_0 \in \integer$ and initial state $x_{t_0}$, where $t \in \integer$ is the time and $x_t \in \real^n$, $u_t \in \real^m$, and $y_t \in \real^p$ are the state, input, and output of the system. The system \eqref{Eq:LTV} is said to be \emph{linear time-periodic (LTP)} if there exists $T \in \natural$ (a \emph{period}) such that $A_{t+T} = A_t$ (and similarly for $B_t, C_t, D_t$) for all $t \in \integer$. The smallest $T$ satisfying this condition is the  \emph{fundamental period}; without loss of generality, we assume going forward that $T$ is the fundamental period. Note that when $T=1$, the system \eqref{Eq:LTV} is linear time-invariant (LTI). A discrete-time LTP model may arise naturally in discrete time, or may have been obtained via appropriate sampling of a continuous-time LTP system.

\begin{subequations}
For the LTV system \eqref{Eq:LTV} and integers $t_1,t_2$ with $t_1 \leq t_2$, the state-transition matrix $\Phi^{t_2}_{t_1} \in \real^{n\times n}$ and impulse response matrix $G^{t_2}_{t_1} \in \real^{p \times m}$ from step $t_1$ to $t_2$ are defined as
\begin{align}
    \label{Eq:SysMatNotation_Phi}
    \Phi^{t_2}_{t_1} :=&\; \small \begin{cases}
        I, &\text{if}\,\, t_2 = t_1\\
        A_{t_2-1} A_{t_2-2} \cdots A_{t_1}, &\text{if}\,\, t_2 > t_1,
    \end{cases} \\ \label{Eq:SysMatNotation_G}
    G^{t_2}_{t_1} :=&\; \small \begin{cases}
    D_{t_1}, &\text{if}\,\, t_2 = t_1\\
        C_{t_2} \Phi^{t_2}_{t_1+1} B_{t_1}, &\text{if}\,\, t_2 > t_1.
    \end{cases}
\end{align}
Similarly, the associated (reversed)  extended controllability matrix $\mathscr{C}^{t_2}_{t_1} \in \real^{n \times (t_2-t_1+1)m}$, the extended observability matrix $\mathscr{O}^{t_2}_{t_1} \in \real^{(t_2-t_1+1)p \times n}$, and the block matrix $\mathscr{I}^{t_2}_{t_1}$ of impulse-response coefficients are defined as
\begin{align}
    \label{Eq:SysMatNotation_C}
    \mathscr{C}^{t_2}_{t_1} :=& \big[
        \Phi^{t_2+1}_{t_1+1} B_{t_1},\;
        \Phi^{t_2+1}_{t_1+2} B_{t_1+1},\;
        \ldots,\;
        \Phi^{t_2+1}_{t_2+1} B_{t_2}
    \big],
    \\ \label{Eq:SysMatNotation_O}
    \mathscr{O}^{t_2}_{t_1} :=& \big[
        C_{t_1} \Phi^{t_1}_{t_1};\;
        C_{t_1+1} \Phi^{t_1+1}_{t_1};\;
        \ldots;\;
        C_{t_2} \Phi^{t_2}_{t_1}
    \big],
    \\ \label{Eq:SysMatNotation_I}
    \mathscr{I}^{t_2}_{t_1} :=& \small \begin{bmatrix}
        G^{t_1}_{t_1} \\
        G^{t_1+1}_{t_1} & G^{t_1+1}_{t_1+1} \\
        \vdots & \vdots & \ddots \\
        G^{t_2}_{t_1} & G^{t_2}_{t_1+1} & \cdots & G^{t_2}_{t_2} \\
    \end{bmatrix}.
\end{align}
With this notation, the unique solution of \eqref{Eq:LTV} with initial condition $x_{t_1}$ at time $t = t_1$ can be expressed as
\begin{align}
    \label{Eq:SysMatNotation:StateSolution}
    x_{t_2} =&\; \Phi^{t_2}_{t_1} x_{t_1} + \mathscr{C}^{t_2-1}_{t_1} u_{[t_1,t_2)},
    \\
    \label{Eq:SysMatNotation:OutputSolution}
    y_{[t_1,t_2]} =&\; \mathscr{O}^{t_2}_{t_1} x_{t_1} + \mathscr{I}^{t_2}_{t_1} u_{[t_1,t_2]},
\end{align}
for any $t_2 > t_1$ in \eqref{Eq:SysMatNotation:StateSolution} and any $t_2 \geq t_1$ in \eqref{Eq:SysMatNotation:OutputSolution}.

Throughout the paper, we let $w_{[t_1,t_2]} := [u_{[t_1,t_2]}; y_{[t_1,t_2]}]$ denote a trajectory of the system \eqref{Eq:LTV}.
\end{subequations}

\subsection{Lifting an LTP System to an LTI System}

We now recall a classical technique for ``lifting'' an LTP system into an LTI system \cite{INTRO:LIFT:Bittanti2000}.

\begin{subequations}
\begin{definition}[Lift of an LTP System]
    For an LTP system $\Sys$ as in \eqref{Eq:LTV} of period $T$ and an initial time $t_0 \in \integer$, the associated \emph{lifted system $\LiftSys(t_0)$ of $\Sys$ with initial time $t_0$} is the LTI system
    \begin{align} \label{Eq:LiftSys}
    \LiftSys(t_0) : \bigg\{ \begin{aligned}
        \liftvec{x}_{\tau+1} =&\; \liftmat{A} \liftvec{x}_\tau + \liftmat{B} \liftvec{u}_\tau \\
        \liftvec{y}_\tau =&\; \liftmat{C} \liftvec{x}_\tau + \liftmat{D} \liftvec{u}_\tau
    \end{aligned} \end{align}
    with state $\liftvec{x}_\tau \in \real^n$, input $\liftvec{u}_{\tau} \in \real^{mT}$, output $\liftvec{y}_{\tau}\in\real^{pT}$, and time $\tau \in\integer$, where
    \begin{equation}\label{Eq:LiftedMatrices}
    \begin{aligned}
        \liftmat{A} &:= \Phi^{t_0+T}_{t_0},\\
        \liftmat{C} &:= \mathscr{O}^{t_0+T-1}_{t_0},
    \end{aligned}
    \qquad
    \begin{aligned}
        \liftmat{B} &:= \mathscr{C}^{t_0+T-1}_{t_0},\\
        \liftmat{D} &:= \mathscr{I}^{t_0+T-1}_{t_0}.
    \end{aligned}
    \end{equation}
\end{definition}
\end{subequations}

\begin{figure}[t]
\centerline{\resizebox{0.5\textwidth}{!}{

\begin{tikzpicture}
    \tikzstyle{box} = [draw, rectangle, minimum height=15pt, minimum width=45pt]
    \tikzstyle{bibox} = [draw, rectangle, minimum height=30pt, minimum width=45pt]
    
    \draw[dashed] (-70pt,7.5pt) -- (250pt,7.5pt);
    \draw[dashed] (-70pt,-52.5pt) -- (250pt,-52.5pt);
    \draw[dashed] (-70pt,-112.5pt) -- (250pt,-112.5pt);
    
    \draw[->] (-55pt,-15pt) -- (-55pt,7.5pt);
    \draw[->] (-55pt,-30pt) -- (-55pt,-52.5pt);
    \node[] at (-55pt,-22.5pt) {period $T$};
    \draw[->] (-55pt,-75pt) -- (-55pt,-52.5pt);
    \draw[->] (-55pt,-90pt) -- (-55pt,-112.5pt);
    \node[] at (-55pt,-82.5pt) {period $T$};
    
    \node[] at (0pt,15pt) {State};
    \node[box, fill=orange!20] at (0pt,0pt) {$x_{t_0}$};
    \node[box, fill=orange!5] at (0pt,-15pt) {$x_{t_0+1}$};
    \node[bibox, fill=orange!5] at (0pt,-37.5pt) {};
    \node[] at (0pt,-32.5pt) {$\vdots$};
    \node[box, fill=orange!20] at (0pt,-60pt) {$x_{t_0+T}$};
    \node[box, fill=orange!5] at (0pt,-75pt) {$x_{t_0+T+1}$};
    \node[bibox, fill=orange!5] at (0pt,-97.5pt) {};
    \node[] at (0pt,-92.5pt) {$\vdots$};
    \node[box, fill=orange!20] at (0pt,-120pt) {$x_{t_0+2T}$};
    \fill [orange!5] (-22.5pt, -127.5pt) rectangle (22.5pt, -150pt);
    \node[] at (0pt,-135pt) {$\vdots$};
    \draw[] (-22.5pt,-127.5pt) -- (-22.5pt,-150pt);
    \draw[] (22.5pt,-127.5pt) -- (22.5pt,-150pt);
    
    \draw[<->] (22.5pt,0pt) -- (37.5pt,0pt);
    \node[] at (47.5pt,0pt) {$\liftvec{x}_0$};
    \draw[<->] (22.5pt,-60pt) -- (37.5pt,-60pt);
    \node[] at (47.5pt,-60pt) {$\liftvec{x}_1$};
    \draw[<->] (22.5pt,-120pt) -- (37.5pt,-120pt);
    \node[] at (47.5pt,-120pt) {$\liftvec{x}_2$};
    
    \node[] at (100pt,15pt) {Input};
    \node[box, fill=green!16] at (100pt,0pt) {$u_{t_0}$};
    \node[box, fill=green!16] at (100pt,-15pt) {$u_{t_0+1}$};
    \node[bibox, fill=green!16] at (100pt,-37.5pt) {};
    \node[] at (100pt,-32.5pt) {$\vdots$};
    \node[box, fill=green!16] at (100pt,-60pt) {$u_{t_0+T}$};
    \node[box, fill=green!16] at (100pt,-75pt) {$u_{t_0+T+1}$};
    \node[bibox, fill=green!16] at (100pt,-97.5pt) {};
    \node[] at (100pt,-92.5pt) {$\vdots$};
    \node[box, fill=green!16] at (100pt,-120pt) {$u_{t_0+2T}$};
    \fill [green!16] (77.5pt, -127.5pt) rectangle (122.5pt, -150pt);
    \node[] at (100pt,-135pt) {$\vdots$};
    \draw[] (77.5pt,-127.5pt) -- (77.5pt,-150pt);
    \draw[] (122.5pt,-127.5pt) -- (122.5pt,-150pt);
    
    \draw [decorate,
    decoration = {brace}] (126.5pt,7.5pt) --  (126.5pt,-52.5pt);
    \draw[<->] (130pt,-22.5pt) -- (145pt,-22.5pt);
    \node[] at (155pt,-22.5pt) {$\liftvec{u}_0$};
    \draw [decorate,
    decoration = {brace}] (126.5pt,-52.5pt) --  (126.5pt,-112.5pt);
    \draw[<->] (130pt,-82.5pt) -- (145pt,-82.5pt);
    \node[] at (155pt,-82.5pt) {$\liftvec{u}_1$};
    
    \node[] at (200pt,15pt) {Output};
    \node[box, fill=blue!8] at (200pt,0pt) {$y_{t_0}$};
    \node[box, fill=blue!8] at (200pt,-15pt) {$y_{t_0+1}$};
    \node[bibox, fill=blue!8] at (200pt,-37.5pt) {};
    \node[] at (200pt,-32.5pt) {$\vdots$};
    \node[box, fill=blue!8] at (200pt,-60pt) {$y_{t_0+T}$};
    \node[box, fill=blue!8] at (200pt,-75pt) {$y_{t_0+T+1}$};
    \node[bibox, fill=blue!8] at (200pt,-97.5pt) {};
    \node[] at (200pt,-92.5pt) {$\vdots$};
    \node[box, fill=blue!8] at (200pt,-120pt) {$y_{t_0+2T}$};
    \fill [blue!8] (177.5pt, -127.5pt) rectangle (222.5pt, -150pt);
    \node[] at (200pt,-135pt) {$\vdots$};
    \draw[] (177.5pt,-127.5pt) -- (177.5pt,-150pt);
    \draw[] (222.5pt,-127.5pt) -- (222.5pt,-150pt);
    
    \draw [decorate,
    decoration = {brace}] (226.5pt,7.5pt) --  (227.5pt,-52.5pt);
    \draw[<->] (230pt,-22.5pt) -- (245pt,-22.5pt);
    \node[] at (255pt,-22.5pt) {$\liftvec{y}_0$};
    \draw [decorate,
    decoration = {brace}] (226.5pt,-52.5pt) --  (227.5pt,-112.5pt);
    \draw[<->] (230pt,-82.5pt) -- (245pt,-82.5pt);
    \node[] at (255pt,-82.5pt) {$\liftvec{y}_1$};
\end{tikzpicture}

}}
\caption{Relationship between the state $x_t$, input $u_t$, output $y_t$ of an LTP system $\Sys$ and the state $\liftvec{x}_\tau$, input $\liftvec{u}_\tau$, output $\liftvec{y}_\tau$ of its lifted system $\LiftSys(t_0)$.}
\label{FIG:Lift}
\end{figure}
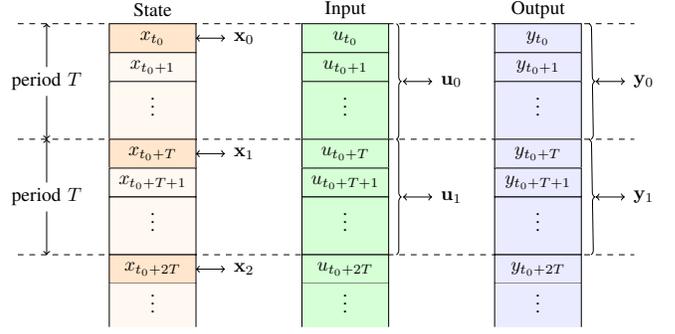

The idea behind lifting is that each time step $\tau$ of the lifted system $\LiftSys(t_0)$ corresponds to $T$ successive time steps of the original LTP system $\Sys$. The state/input/output of $\LiftSys(t_0)$ are related to the state/input/output of $\Sys$ via
\begin{align*}
    \liftvec{x}_\tau =&\; x_{t_0 + \tau T}, \\
    \liftvec{u}_\tau =&\; u_{[t_0 + \tau T,\, t_0 + (\tau+1)T)}, \\
    \liftvec{y}_\tau =&\; y_{[t_0 + \tau T,\, t_0 + (\tau+1)T)}.
\end{align*}
Each input vector $\liftvec{u}_\tau$ (or output vector $\liftvec{y}_\tau$) of $\LiftSys(t_0)$ stacks the inputs (or outputs) of $\Sys$ over one period,
and the state vector $\liftvec{x}_\tau$ is the state of $\Sys$ at the ``beginning'' of this period, as specified by the initial time $t_0$; see Fig. \ref{FIG:Lift}. Note from \eqref{Eq:LiftedMatrices} that the matrices $\liftmat{A}, \liftmat{B}, \liftmat{C}, \liftmat{D}$ depend on the initial time $t_0$. Nonetheless, some properties of the lifted system --- such as the eigenvalues of $\liftmat{A}$ --- are invariant under the choice of the initial time $t_0$. See \cite{LIFT:Jemaa2003} for more information on lifting and properties of the lifted system.

\section{Behavioral Systems Theory for Linear Time-Periodic Systems} \label{SEC:Beh}

In this section we develop a set of results on behavioral systems theory for linear time-periodic systems.

\subsection{Behavioural Representation of LTV Systems}

In the framework of behavioral systems theory, the input-output trajectories of the system \eqref{Eq:LTV} are described independent of the state representation through the \emph{behavior}.

\begin{definition}[Behavior] \label{DEF:Beh}
    For the LTV system $\Sys$ in \eqref{Eq:LTV} and an integer $t_1$, the \emph{behavior}  $\mathscr{B}^\Sys_{[t_1,\infty)}$ of $\mathcal{S}$ on the time interval $[t_1, \infty) \cap \mathbb{Z}$ is the set
    \begin{align*}
        \mathscr{B}^\Sys_{[t_1,\infty)} :=& \left\{
        {\footnotesize \begin{bmatrix}
            u_{[t_1,\infty)} \\ y_{[t_1,\infty)}
        \end{bmatrix}}\,\,
        \middle|\,\,
        \exists\, x_{t_1}\; \text{s.t.}\; \eqref{Eq:LTV}\; \text{holds for all}\; t \geq t_1
        \right\} .
        \end{align*}
\end{definition}

Given $\mathscr{B}^\Sys_{[t_1,\infty)}$ and an integer $t_2 \geq t_1$, we let $\mathscr{B}^\Sys_{[t_1,t_2]}$ denote the restriction of $\mathscr{B}^\Sys_{[t_1,\infty)}$ to the interval $[t_1,t_2]$, and in the case $t_2 > t_1$, we let $\mathscr{B}^\Sys_{[t_1,t_2)} := \mathscr{B}^\Sys_{[t_1,t_2-1]}$. 
The behavior defines a subspace of the vector space of semi-infinite sequences, and contains all possible input-output trajectories of the system. 
Going forward, we focus primarily on the restricted behavior.

\begin{lemma} \label{LEMMA:BehColSpan}
    The restricted behavior  $\mathscr{B}^\Sys_{[t_1,t_2]}$ of the LTV system $\mathcal{S}$ in \eqref{Eq:LTV} is a finite-dimensional vector space 
    and
    \begin{align*}
        \mathscr{B}^\Sys_{[t_1,t_2]} =&\; \mathrm{ColSpan} \small \begin{bmatrix} 0 & I \\ \mathscr{O}^{t_2}_{t_1} & \mathscr{I}^{t_2}_{t_1} \end{bmatrix}.
    \end{align*}
\end{lemma}

\begin{corollary} \label{COROLLARY:BehDim}
    $\dim \mathscr{B}^\Sys_{[t_1,t_2]} = \rank(\mathscr{O}^{t_2}_{t_1}) + m(t_2-t_1+1)$.
\end{corollary}

When $\Sys$ is an LTI system, the behavior is invariant under shift of the time interval, meaning that
\begin{align*}
    \mathscr{B}^\Sys_{[t_1, t_2]} = \mathscr{B}^\Sys_{[t_1 + s, t_2 + s]} 
    \quad \text{for all } s \in\integer.
\end{align*}
This follows from Definition \ref{DEF:Beh}, due to shift-invariance of the system matrices. A similar result holds when $\Sys$ is an LTP system of period $T$, namely that
\begin{align*}
    \mathscr{B}^\Sys_{[t_1, t_2]} = \mathscr{B}^\Sys_{[t_1 + s T, t_2 + s T]}
    \quad \text{for all } s \in\integer.
\end{align*}
Unlike the case of LTI or LTP systems, the behavior of a general LTV system is not a shift-invariant subspace. However, given the behavior over an interval, the behavior on the first several steps can be easily constructed.

\begin{lemma} \label{LEMMA:SubBehRepre}
    For the LTV system $\Sys$ in \eqref{Eq:LTV} and integers $t_0 \leq t_1 \leq t_2$, if 
    \begin{align*}
        \mathscr{B}^\Sys_{[t_0, t_2]}
        = \mathrm{ColSpan} [ U_{t_0}; \ldots; U_{t_2}; Y_{t_0}; \ldots; Y_{t_2} ]
    \end{align*}
    for some matrices $U_{t_0}, \ldots, U_{t_2} \in \real^{m \times h}$ and $Y_{t_0}, \ldots, Y_{t_2} \in \real^{p \times h}$ with some $h \in \natural$, then
    \begin{align*}
        \mathscr{B}^\Sys_{[t_0, t_1]}
        = \mathrm{ColSpan} [ U_{t_0}; \ldots; U_{t_1}; Y_{t_0}; \ldots; Y_{t_1} ].
    \end{align*}
\end{lemma}

Different state-space models may correspond to a same behavior.
The following result characterizes when different LTV systems have the same restricted behavior.

\begin{lemma} \label{LEMMA:DiffModelSameBeh}
    \begin{subequations}
    For LTV systems $\Sys, \bar \Sys$ and integers $t_1 \leq t_2$, 
    we have $\mathscr{B}^\Sys_{[t_1,t_2]} = \mathscr{B}^{\bar \Sys}_{[t_1,t_2]}$ if, and only if,
    \begin{align}&
        \label{Eq:LEMMA:DiffModelSameBeh:Cond1}
        \mathcal{O} := \mathrm{ColSpan}(\mathscr{O}^{t_2}_{t_1}) = \mathrm{ColSpan}(\mathscr{\bar O}^{t_2}_{t_1}),
        \\& \label{Eq:LEMMA:DiffModelSameBeh:Cond2}
        \mathrm{ColSpan}(\mathscr{I}^{t_2}_{t_1} - \mathscr{\bar I}^{t_2}_{t_1}) \subseteq \mathcal{O},
    \end{align}
    where $\mathscr{O}^{t_2}_{t_1}$ and $\mathscr{I}^{t_2}_{t_1}$ (resp. $\mathscr{\bar O}^{t_2}_{t_1}$ and $\mathscr{\bar I}^{t_2}_{t_1}$) are defined as in \eqref{Eq:SysMatNotation_O} and \eqref{Eq:SysMatNotation_I} for system $\Sys$ (resp. $\bar\Sys$).
    \end{subequations}
\end{lemma}

\subsubsection{Controllability}
In the behavioral framework, controllability is defined in a trajectory-based sense,
as opposed to the more classical notion of state-controllability.

\begin{definition}[Controllability \cite{BEHAV:Markovsky2021, Ilchmann2005}] \label{DEF:Ctrb}
    An LTV system $\Sys$ is \emph{controllable} if for any $t_0 \in \integer$, any two trajectories $w^\text{I}_{[t_0,\infty)}$, $w^\text{II}_{[t_0,\infty)} \in \mathscr{B}^\Sys_{[t_0,\infty)}$, and any time $t_1 \geq t_0$,
    there exists a time $t_2 \geq t_1$ and a trajectory $w^\diamond_{[t_0,\infty)} \in \mathscr{B}^\Sys_{[t_0,\infty)}$ such that
    \begin{align} \label{Eq:DEF:Ctrb}
        w^\diamond_{[t_0,t_1)} 
        = w^\text{I}_{[t_0,t_1)}, \qquad
        w^\diamond_{[t_2,\infty)}
        = w^\text{II}_{[t_2,\infty)}.
    \end{align}
\end{definition}

Put differently, an LTV system is controllable if we can ``drive'' from one trajectory to any other trajectory in a finite number of time steps. When $\Sys$ is an LTI system, the second equality in \eqref{Eq:DEF:Ctrb} is sometimes replaced by $w^\diamond_{[t_2,\infty)} = w^\text{II}_{[t_0,\infty)}$ in the literature (e.g. \cite{BEHAV:Willems2005, Coulson2019}).
This alternative definition is equivalent to Definition \ref{DEF:Ctrb} for LTI systems \cite[Remark 4(i)]{Ilchmann2005}.

\subsection{A Definition of Order and Lag for LTV Systems}

\begin{subequations}

In behavioral systems theory, the \emph{order} and \emph{lag} are so-called integer invariants of an LTI system, and can be expressed using a minimal state representation of the behavior. In this subsection, we generalize those notions to LTV systems, and do so in a manner that avoids introducing notions of minimality for LTV models.

First, we review the LTI definitions of order and lag from the literature (e.g., \cite{BEHAV:Markovsky2020, BEHAV:Markovsky2021}).
For an LTI state-space model $\Sys:(A,B,C,D)$, a \emph{minimal representation} $\bar\Sys$ of its behavior $\mathscr{B}^\Sys_{[0,\infty)}$ is a state-space model $\bar\Sys:(\bar A, \bar B, \bar C, \bar D)$, having the minimal possible state dimension and sharing the same behavior as of $\Sys$, i.e., $\mathscr{B}^\Sys_{[0,\infty)} = \mathscr{B}^{\bar\Sys}_{[0,\infty)}$.
The order $\order(\Sys)$ of $\Sys$ is equal to the state dimension of $\bar\Sys$.
Define matrix $\mathscr{O}^{t_2}_{t_1}$ (resp. $\mathscr{\bar O}^{t_2}_{t_1}$) from \eqref{Eq:SysMatNotation_O} for system $\Sys$ (resp. $\bar\Sys$).
The extended observability matrix $\mathscr{\bar O}^{s-1}_0$ of $\bar\Sys$ reaches full column rank equal to $\order(\Sys)$ when $s$ is sufficiently large,
and the lag $\lag(\Sys)$ of $\Sys$ is the smallest integer $s$ that $\mathscr{\bar O}^{s-1}_0$ has full column rank.
We can express both $\order(\Sys)$ and $\lag(\Sys)$ in terms of the model $\Sys$, which is not necessarily a minimal representation, as
\begin{align} \label{Eq:OrderLag_LTI} \begin{aligned}
    \order(\Sys) =&\; \lim_{s \to \infty} \rank (\mathscr{\bar O}^{s-1}_0) 
    = \lim_{s \to \infty} \rank (\mathscr{O}^{s-1}_0),
    \\
    \lag(\Sys) =&\; \min \{ s \in \natural : \rank(\mathscr{\bar O}^{s-1}_0) = \order(\Sys) \}
    \\ =&\; \min \{ s \in \natural : \rank(\mathscr{O}^{s-1}_0) = \order(\Sys) \},
\end{aligned} \end{align}
where we used the equality $\rank(\mathscr{O}^{s-1}_0) = \rank(\mathscr{\bar O}^{s-1}_0)$ which follows from $\mathscr{B}^\Sys_{[0,s)} = \mathscr{B}^{\bar\Sys}_{[0,s)}$ and Corollary \ref{COROLLARY:BehDim}.

Motivated by the above considerations, we introduce the following definition.

\begin{definition}[\bf Order and Lag] \label{DEF:OrderLag}
For the LTV system $\Sys$ in \eqref{Eq:LTV}, the \emph{order} $\order(\Sys,t)$ at time $t$ and \emph{lag} $\lag(\Sys,t)$ at time $t$ are\footnote{$\order(\Sys,t)$ is well-defined in \eqref{Eq:OrderLag_LTV}, since $\rank(\mathscr{O}^{t+s-1}_{t})$ is bounded by the state dimension $n$ and is non-decreasing as we increase $s$ because $\mathscr{O}^{t+s-1}_{t}$ is augmented with extra rows.
Thus, $\lag(\Sys,t)$ is also well-defined in \eqref{Eq:OrderLag_LTV}.}
\begin{align} \label{Eq:OrderLag_LTV} \begin{aligned}
    \order(\Sys, t) :=&\; \lim_{s\to\infty} \rank(\mathscr{O}^{t+s-1}_{t}), 
    \\
    \lag(\Sys, t) :=&\; \min 
    \{ s \in \natural : \rank(\mathscr{O}^{t+s-1}_{t}) = \order(\Sys, t) \}.
\end{aligned} 
\end{align}
\end{definition}

When $\Sys$ is an LTI system, we write its order and lag in the sense of \eqref{Eq:OrderLag_LTV} as $\order(\Sys)$ and $\lag(\Sys)$ respectively, since they are time-independent. 
The definitions in \eqref{Eq:OrderLag_LTV} are consistent with \eqref{Eq:OrderLag_LTI} in the LTI case.
Moreover, via Corollary \ref{COROLLARY:BehDim},
\begin{align} \label{Eq:LagBehDim}
    \dim \mathscr{B}^\Sys_{[t,t+L)}
    = \order(\Sys,t) + m L
    \qquad \forall t \in \mathbb{Z}
\end{align}
for all integers $L \geq \lag(\Sys,t)$, which coincides with an established result \cite[Cor. 5]{BEHAV:Markovsky2020} or \cite[Eq. (1)]{BEHAV:Markovsky2021} in the LTI case.
\end{subequations}

The lag specifies a sufficient length of a trajectory such that, with any subsequent input, the resulting output after the trajectory is uniquely determined, as captured in Lemma \ref{LEMMA:LagIniCond}(ii).
This result generalizes \cite[Lemma 1]{Markovsky2008} or \cite[Lemma 1]{BEHAV:Markovsky2021} which is for the LTI case.
Lemma \ref{LEMMA:LagIniCond}(iii) gives an expression for the unique output, generalizing \cite[Lemma 2]{Fiedler2021} as the LTI case.

\begin{lemma}[Uniqueness of Future Output] \label{LEMMA:LagIniCond}
\begin{subequations}
    Consider the LTV system $\Sys$ in \eqref{Eq:LTV}, a time step $t \in \integer$, and positive integers $L, N$. The following statements hold:
\begin{enumerate}[(i)]
    \item For any trajectory $w_{[t-L,t)} \in \mathscr{B}^\Sys_{[t-L, t)}$ and any input $u^*_{[t,t+N)}$,
    there exists an output $y^*_{[t,t+N)}$ satisfying
    \begin{align} \label{Eq:LEMMA:LagIniCond:Beh}
        \!\!\!\!\!\!\!\!\!
        [u_{[t-L,t)}; u^*_{[t,t+N)}; y_{[t-L,t)}; y^*_{[t,t+N)}]
        \in \mathscr{B}^\Sys_{[t-L,t+N)}.
        \!\!\!
    \end{align}
    \item If $L \geq \lag(\Sys, t-L)$, the output $y^*_{[t,t+N)}$ from (i) is unique.
    \item Moreover, if the behavior $\mathscr{B}^\Sys_{[t-L,t+N)}$ can be expressed as \vspace{-.8em}
    \begin{align} \label{Eq:LEMMA:LagIniCond:Span}
        \mathscr{B}^\Sys_{[t-L,t+N)} 
        = \mathrm{ColSpan} \, [U_{\rm p}; U_{\rm f}; Y_{\rm p}; Y_{\rm f}]
    \end{align}
    for some matrices $U_{\rm p} \in \real^{mL\times h}, U_{\rm f} \in \real^{mN\times h}, Y_{\rm p} \in \real^{pL\times h}, Y_{\rm f} \in \real^{pN\times h}$ with some $h \in \natural$, 
    then the unique output $y^*_{[t,t+N)}$ from (ii) is given as\footnote{Note that $y^*_{[t,t+N)}$ is unique even though the matrices $U_{\rm p}, U_{\rm f}, Y_{\rm p}, Y_{\rm f}$ may not be unique.} \vspace{-.5em}
    \begin{align} \label{Eq:LEMMA:LagIniCond:Repre}
        y^*_{[t,t+N)} = Y_{\rm f}
        \small \begin{bmatrix}
            U_{\rm p} \\ U_{\rm f} \\ Y_{\rm p}
        \end{bmatrix}^\dagger
        \begin{bmatrix} 
            u_{[t-L,t)} \\ u^*_{[t,t+N)} \\ y_{[t-L,t)}
        \end{bmatrix}.
    \end{align}
\end{enumerate}
\end{subequations}
\end{lemma}

\subsection{Behavioral Systems Theory for LTP Systems}

Now we limit our discussion to LTP systems.
We first establish the relationship between the behavior of an LTP system and the behavior of any corresponding lifted system.

\begin{lemma} \label{LEMMA:LiftEqv}
    For an LTP system $\Sys$ of period $T$ and its lifted system $\LiftSys(t_0)$ with initial step $t_0 \in \integer$, it holds that
    \begin{align*}
        \mathscr{B}^\Sys_{[t_0,\, t_0 + s T)} =
        \mathscr{B}^{\LiftSys(t_0)}_{[0, s)}
        \qquad \forall s \in \natural.
    \end{align*}
\end{lemma}

\begin{remark}[\bf Dependence on Initial Step $t_0$]
The lifted system and its behavior depend on the initial step $t_0$.
For instance, consider the following single-state SISO LTP system $\Sys$ of period $T=2$.
\begin{align*} \Sys: \small \bigg\{ \begin{aligned}
    x_{t+1} =&\; x_t + (-1)^t u_t\\
    y_t =&\; x_t
\end{aligned} \end{align*}
The corresponding lifted system $\LiftSys(t_0)$ for $t_0 \in \integer$.
\begin{align*} 
    \LiftSys(t_0): \small \bigg\{ 
\begin{aligned}
    \liftvec{x}_{\tau+1} 
    =&\; \liftvec{x}_\tau 
    + (-1)^{t_0} \begin{bsmallmatrix} 1 & -1 \end{bsmallmatrix} \liftvec{u}_\tau\\
    \liftvec{y}_\tau 
    =&\; \begin{bsmallmatrix} 1 \\ 1 \end{bsmallmatrix} \liftvec{x}_\tau 
    + (-1)^{t_0} \begin{bsmallmatrix} 0 & 0 \\ 1 & 0 \end{bsmallmatrix} \liftvec{u}_\tau
\end{aligned}
\end{align*}
It follows (via Lemma \ref{LEMMA:BehColSpan}) that for $t_0 \in \integer$ the restricted behavior of $\LiftSys(t_0)$ on interval $[0,0]$ is
\begin{align*}
    \mathscr{B}^{\LiftSys(t_0)}_{[0,0]} 
    = \mathrm{ColSpan} \scriptsize \left[\begin{array}{c|cc}
        0 & 1 & 0 \\
        0 & 0 & 1 \\ \hline
        1 & 0 & 0 \\
        1 & (-1)^{t_0} & 0
    \end{array}\right].
\end{align*}
One can now observe that  $\mathscr{B}^{\LiftSys(0)}_{[0,0]}$ and $\mathscr{B}^{\LiftSys(1)}_{[0,0]}$ are different subspaces.
Hence, it is necessary to specify the initial time $t_0$ when introducing the lifted system.
\oprocend
\end{remark}

\subsubsection{Order and Lag}
Notions of order and lag for LTV systems have been introduced in Definition \ref{DEF:OrderLag}.
The next result relates the order and lag of an LTP system to the order and lag of its lifted system.

\begin{lemma} \label{LEMMA:OrderLagLift_LTP}
    For an LTP system $\Sys$ of period $T$, we have (i) $\order(\LiftSys(t)) = \order(\Sys,t)$, and (ii) $\lag(\LiftSys(t)) = \lceil \lag(\Sys, t) / T \rceil$.
\end{lemma}

For unknown LTP systems with known periods and state dimensions, we can establish bounds of their orders and lags.

\begin{corollary} \label{COROLLARY:OrderLagBound_LTP}
    For an LTP system $\Sys$ as in \eqref{Eq:LTV} of period $T$,
    we have (i) $\order(\Sys, t) \leq n$, and (ii) $\lag(\Sys, t) \leq nT$.
\end{corollary}

\subsubsection{Controllability}
The controllability of an LTP system is equivalent to the controllability of its lifted systems.

\begin{lemma} \label{LEMMA:CtrbEqv}
    An LTP system $\Sys$ is controllable if, and only if, its lifted systems $\LiftSys(t_0)$ are controllable for all $t_0 \in \integer$.
\end{lemma}

\subsection{A Fundamental Lemma for LTP Systems}

According to the so-called Fundamental Lemma \cite[Thm. 1]{BEHAV:Willems2005}, under technical conditions, the restricted behavior of an LTI system can be completely described via recorded historical data. This result is reviewed as Lemma \ref{LEMMA:Fund} below. We first review the notion of persistent excitation.

\begin{definition}[Persistent Excitation]
    A sequence $z_{[t_1, t_2]}$ is \emph{persistently exciting (p.e.) of order $K$}, for positive integer $K \leq t_2-t_1+1$, if the associated block-Hankel matrix of depth $K$ \vspace{-.5em}
    \begin{align*} \footnotesize
        \mathcal{H}_K(z_{[t_1, t_2]})
        := \begin{bmatrix}
        z_{t_1} & z_{t_1+1} & \cdots & z_{t_2-K+1} \\
        z_{t_1+1} & z_{t_1+2} & \cdots & z_{t_2-K+2} \\[-.3em]
        \vdots & \vdots & \ddots & \vdots \\[-.3em]
        z_{t_1+K-1} & z_{t_1+K} & \cdots & z_{t_2} \\
        \end{bmatrix}
    \end{align*}
    has full row rank.
\end{definition}

\begin{lemma}[Fundamental Lemma \cite{BEHAV:Willems2005}] \label{LEMMA:Fund}
    Let $\Sys$ be an LTI system, and
    let $w^{\rm d}_{[t_1, t_2]}$ be a trajectory of $\Sys$. For $K \in \natural$, if 
    \begin{enumerate}[(i)]
    \item $\Sys$ is controllable, and
    \item $u^{\rm d}_{[t_1, t_2]}$ is p.e. of order $K+\order(\Sys)$,
    \end{enumerate}
    then \vspace{-.8em}
    \begin{align*}
        \mathrm{ColSpan} \big( \mathcal{H}_K(w^{\rm d}_{[t_1, t_2]}) \big)
        = \mathscr{B}^\Sys_{[0, K)},
    \end{align*}
    where $\mathcal{H}_K(w^{\rm d}_{[t_1, t_2]}) := \big[ \mathcal{H}_K(u^{\rm d}_{[t_1, t_2]}); \mathcal{H}_K(y^{\rm d}_{[t_1, t_2]}) \big]$.
\end{lemma}


Based on the lifting operation, we now define a natural extension of persistent excitation for LTP systems, and present a corresponding version of the fundamental lemma.

\begin{definition}[Periodic Persistent Excitation]
    A sequence $z_{[t_1, t_2]}$ is \emph{$T$-periodically persistently exciting ($T$-p.p.e.) of order $K$}, for $K,T\in\natural$ satisfying $K \leq t_2-t_1+1$, if
    \begin{align*} \footnotesize
        \mathcal{H}^T_K(z_{[t_1, t_2]}) 
        := \begin{bmatrix}
            z_{t_1} & z_{t_1+T} & \cdots & z_{t_1+PT} \\
            z_{t_1+1} & z_{t_1+T+1} & \cdots & z_{t_1+PT+1} \\[-.3em]
            \vdots & \vdots & \ddots & \vdots \\[-.3em]
            z_{t_1+K-1} & z_{t_1+T+K-1} & \cdots & z_{t_1+PT+K-1} \\
        \end{bmatrix}
    \end{align*}
    has full row rank, 
    where $P := \lfloor (t_2-t_1-K+1)/T \rfloor$.\footnote{One can observe that $\mathcal{H}^T_K (z_{[t_1, t_2]})$ is obtained by retaining every $T$-th column of $\mathcal{H}_K (z_{[t_1, t_2]})$.}
\end{definition}

\begin{lemma}[Fundamental Lemma for LTP Systems] \label{LEMMA:Fund_LTP}
    Let $\Sys$ be an LTP system of period $T$, and let $w^{\rm d}_{[t_1, t_2]}$ be a trajectory of $\Sys$ on interval $[t_1, t_2]$.
    For $K \in \natural$, if 
    \begin{enumerate}[(i)]
    \item $\Sys$ is controllable, and
    \item $u^{\rm d}_{[t_1, t_2]}$ is $T$-p.p.e. of order $(\lceil K/T \rceil + \order(\Sys,t_1)) T$,
    \end{enumerate}
    then \vspace{-.8em}
    \begin{align} \label{Eq:LEMMA:Fund_LTP}
        \mathrm{ColSpan} \big( \mathcal{H}^T_K(w^{\rm d}_{[t_1, t_2]}) \big)
        = \mathscr{B}^\Sys_{[t_1, t_1 + K)},
    \end{align}
    where $\mathcal{H}^T_K(w^{\rm d}_{[t_1, t_2]}) := \big[ \mathcal{H}^T_K(u^{\rm d}_{[t_1, t_2]}); \mathcal{H}^T_K(y^{\rm d}_{[t_1, t_2]}) \big]$
\end{lemma}

When $\order(\Sys, t_1)$ is unknown but bounded by some $n \in\integer$, we may obtain (ii) in Lemma \ref{LEMMA:Fund_LTP} by requiring the input $u^{\rm d}_{[t_1,t_2]}$ to be $T$-p.p.e. of a sufficient order $(\lceil K/T \rceil + n) T$.
This is because by definition a signal being $T$-p.p.e of order $K'$ is also $T$-p.p.e. of any smaller order $K''\leq K'$.
\setlength{\abovedisplayskip}{.5em}
\setlength{\belowdisplayskip}{.5em}

\section{Data-Driven Model Predictive Control for Linear Time-Periodic Systems} \label{SEC:Alg}
\label{Sec:DDMPC}

Based on our previous results extending behavioral systems theory to LTP systems, in this section, we develop a DDMPC algorithm for LTP systems $\Sys$ as in \eqref{Eq:LTV} of known period $T$.

\subsection{Prediction, Control, and Initial Horizons}

We consider a receding-horizon control strategy, in which at time $t$ the control signal $u$ on interval $[t, t+N_{\rm c}) \cap\integer$ (the \emph{control horizon}) is computed by minimizing an appropriate cost function of the predicted trajectory over a finite horizon $[t, t+N) \cap \mathbb{Z}$ (the \emph{prediction horizon}), where $N_{\rm c}, N \in \natural$ are design parameters with $N_{\rm c} \leq N$. 

In the present data-driven scenario, the initial condition of the system at time $t$ is specified by the recent trajectory in a past interval $[t-L, t) \cap \mathbb{Z}$ called the \emph{initial horizon}, with parameter $L \in \natural$. According to Lemma \ref{LEMMA:LagIniCond}, if $L \geq \lag(\Sys, t-L)$, we can uniquely predict the future output, given any future input. Notice via Corollary \ref{COROLLARY:OrderLagBound_LTP} that the lag $\lag(\Sys, t-L)$ is bounded by $nT$, so the output prediction is always unique when we select $L \geq nT$.
We call the union $[t-L,t+N) \cap\integer$ of the initial and prediction horizons as the \emph{total horizon}; see Fig. \ref{FIG4:Hori}.

\subsection{Offline Data Collection}
\label{SUBSEC:Algo:OfflineData}

\begin{subequations}

The restricted behavior $\mathscr{B}^\Sys_{[t-L,t+N)}$ on the total horizon must be known for us to predict future trajectories and compute control actions in the DDMPC framework. 
In previous work on DDMPC for LTI systems \cite{INTRO:DDMPC:Berberich2020a, INTRO:DDMPC:Berberich2020b, INTRO:DDMPC:Berberich2021, Coulson2019, INTRO:Coulson2019b, INTRO:Coulson2021, INTRO:Huang2021a, INTRO:Huang2021b}, the behavior $\mathscr{B}^\Sys_{[t-L,t+N)}$ can be represented using recorded offline data. We may extend this strategy to the case where $\Sys$ is an LTP system.
However, since the system is periodic, its behavior $\mathscr{B}^\Sys_{[t-L,t+N)}$ can equal one of $T$ different possible subspaces, depending on the time $t$. Fortunately, all $T$ possibilities for the behavior $\mathscr{B}^\Sys_{[t-L,t+N)}$ can be covered using collected data.

\subsubsection{\bf Offline Data}
Let $w^{\rm d}_{[t_{\rm d1}, t_{\rm d2}]}$ be offline data collected from the system $\Sys$ on the interval $[t_{\rm d1}, t_{\rm d2}]$, where we require that the input signal $u^{\rm d}_{[t_{\rm d1}, t_{\rm d2}]}$ is $T$-p.p.e. of order $(\lceil K/T \rceil + n) T$, with $K := L+N+T-1$.
Arrange the data into the ``uncropped'' data matrices $U^{\rm d} \in \real^{mK\times h}$ and $Y^{\rm d} \in \real^{pK\times h}$:
\begin{align*}
    U^{\rm d} := \mathcal{H}^T_K(u^{\rm d}_{[t_{\rm d1}, t_{\rm d2}]}), \qquad
    Y^{\rm d} := \mathcal{H}^T_K(y^{\rm d}_{[t_{\rm d1}, t_{\rm d2}]}),
\end{align*}
where $h$ denotes the common width of $U^{\rm d}$ and $Y^{\rm d}$, given by $h := \lfloor (t_{\rm d2} -t_{\rm d1} - K + 1) /T \rfloor + 1$.
We extract from $U^{\rm d}$ and $Y^{\rm d}$ the $T$ sets of \emph{data matrices} $U_{\rm p}^\Idx \in \real^{mL\times h}$, $U_{\rm f}^\Idx \in \real^{mN\times h}$, $Y_{\rm p}^\Idx \in \real^{pL\times h}$ and $Y_{\rm f}^\Idx \in \real^{pN\times h}$, defined as
\begin{align} \label{Eq:DataMatDef}
\begin{aligned}
    U_{\rm p}^\Idx :=&\; U^{\rm d}_{[\Idx, \Idx+L-1]},
    \\
    Y_{\rm p}^\Idx :=&\; Y^{\rm d}_{[\Idx, \Idx+L-1]},
\end{aligned} \qquad
\begin{aligned}
    U_{\rm f}^\Idx :=&\; U^{\rm d}_{[\Idx+L, \Idx+L+N-1]},
    \\
    Y_{\rm f}^\Idx :=&\; Y^{\rm d}_{[\Idx+L, \Idx+L+N-1]},
\end{aligned}
\end{align}
where each set has an exclusive index $\Idx \in \{1, \ldots, T\}$.
In \eqref{Eq:DataMatDef}, we let $U^{\rm d}_{[r_1,r_2]} \in \real^{m(r_2-r_1+1)\times h}$ denote the sub-matrix consisting of the $r_1$-th, ..., $r_2$-th block rows of $U^{\rm d}$, and similarly for $Y^{\rm d}_{[r_1,r_2]}$, with abuse of notation.

\subsubsection{\bf Representation of Behavior}
The matrices $U_{\rm p}^\Idx$, $U_{\rm f}^\Idx$, $Y_{\rm p}^\Idx$, $Y_{\rm f}^\Idx$ built from offline data can represent the behavior on the total horizon at time $t^\Idx := t_{\rm d1} + \Idx + L-1$, as said in the following lemma; see Fig. \ref{FIG4:MatSpan}.

\begin{lemma} \label{LEMMA:DataSpan}
    Consider an LTP system $\Sys$ as in \eqref{Eq:LTV} of period $T$. 
    For $L, N \in \natural$, construct data matrices $U_{\rm p}^\Idx, U_{\rm f}^\Idx, Y_{\rm p}^\Idx, Y_{\rm f}^\Idx$ from \eqref{Eq:DataMatDef} with data $w^{\rm d}_{[t_{\rm d1}, t_{\rm d2}]}$. 
    If $\Sys$ is controllable and $u^{\rm d}_{[t_{\rm d1}, t_{\rm d2}]}$ is $T$-p.p.e. of order $(\lceil K/T \rceil + \order(\Sys, t_{\rm d1})) T$ with $K := L+N+T-1$, 
    then for $\Idx \in \{1, \ldots, T\}$ we have
    \begin{align} \label{Eq:LEMMA:DataSpan}
        \mathrm{ColSpan} \big[ U_{\rm p}^\Idx; U_{\rm f}^\Idx; Y_{\rm p}^\Idx; Y_{\rm f}^\Idx \big] = \mathscr{B}^\Sys_{[t^\Idx-L, t^\Idx+N)}.
    \end{align}
\end{lemma}

Since $\{t^\Idx\}_{\Idx=1}^T$ are consecutive time steps in one period, by periodicity of $\Sys$, the subspaces $\mathscr{B}^\Sys_{[t^\Idx-L, t^\Idx+N)}$ with different selections of the index $\Idx \in \{1,\ldots,T\}$ cover all $T$ possibilities of the behavior $\mathscr{B}^\Sys_{[t-L, t+N)}$ for different time steps $t$.
Define the \emph{proper index} $\ProperIdx(t)$ at time $t$.
\begin{align} \label{Eq:ProperIdxDef}
    \ProperIdx(t) := 1 + (t - t_{\rm d1} - L \mod T)
\end{align}
Thus, $\Idx = \ProperIdx(t)$ is the ``correct'' index $\Idx$ such that the data matrices $U_{\rm p}^\Idx$, $U_{\rm f}^\Idx$, $Y_{\rm p}^\Idx$, $Y_{\rm f}^\Idx$ represent the behavior on the total horizon at time $t$, i.e., 
\begin{align} \label{Eq:OfflineData_OnlineBeh}
    \!\!\!
    \mathrm{ColSpan} \big[ U_{\rm p}^{\ProperIdx(t)}; U_{\rm f}^{\ProperIdx(t)}; Y_{\rm p}^{\ProperIdx(t)}; Y_{\rm f}^{\ProperIdx(t)} \big] = \mathscr{B}^\Sys_{[t-L, t+N)},
    \!\!\!
\end{align}
because of \eqref{Eq:LEMMA:DataSpan}, periodicity of $\Sys$ and the fact that $t - t^{\ProperIdx(t)}$ is a multiple of $T$.
\end{subequations}

\begin{figure}[t]
\centerline{\resizebox{0.5\textwidth}{!}{

\begin{tikzpicture}
    \tikzstyle{grid} = [draw, rectangle, minimum height=10pt, minimum width=10pt]
    
    \fill [orange!7] (0pt, -45pt) rectangle (-140pt, 27pt);
    \fill [blue!4] (0pt, -45pt) rectangle (200pt, 27pt);
    \draw[dashed] (0pt, 5pt) -- (0pt, 27pt);
    
    \node[align=center] at (-120pt, -35pt) {\color{orange!75!black} past};
    \node[align=center] at (180pt, -35pt) {\color{blue} future};
    
    \node[grid, fill=orange!40] at (-5pt, 0pt) {};
    \node[grid, fill=orange!40] at (-15pt, 0pt) {};
    \node[grid, minimum width=50pt, fill=orange!40] at (-45pt, 0pt) {};
    \node[] at (-45pt, 0pt) {$\cdots$};
    \node[grid, fill=orange!40] at (-75pt, 0pt) {};
    \node[grid, fill=orange!40] at (-85pt, 0pt) {};
    \node[grid] at (-95pt, 0pt) {};
    \node[grid] at (-105pt, 0pt) {};
    \node[] at (-120pt, 0pt) {$\cdots$};
    \draw[] (-110pt, 5pt) -- (-125pt, 5pt);
    
    \node[grid, fill=blue!20] at (5pt, 0pt) {};
    \node[grid, minimum width=30pt, fill=blue!20] at (25pt, 0pt) {};
    \node[] at (25pt, 0pt) {$\cdots$};
    \node[grid, fill=blue!20] at (45pt, 0pt) {};
    \node[grid, fill=blue!20] at (55pt, 0pt) {};
    \node[grid, fill=blue!20] at (65pt, 0pt) {};
    \node[grid, minimum width=30pt, fill=blue!20] at (85pt, 0pt) {};
    \node[] at (85pt, 0pt) {$\cdots$};
    \node[grid, fill=blue!20] at (105pt, 0pt) {};
    \node[grid, fill=blue!20] at (115pt, 0pt) {};
    \node[grid] at (125pt, 0pt) {};
    \node[grid] at (135pt, 0pt) {};
    \node[] at (150pt, 0pt) {$\cdots$};
    \draw[] (140pt, 5pt) -- (155pt, 5pt);
    
    \draw[->] (-135pt,-5pt) -- (170pt, -5pt);
    \node[] at (177.5pt, -12.5pt) {time step};
    
    \node[grid, minimum width=50pt, ultra thick] at (25pt, 0pt) {};
    
    \draw (-90pt, -10pt) -- (-90pt, -45pt);
    \draw[<-] (-90pt, -30pt) -- (-77.5pt, -30pt);
    \draw[->] (-12.5pt, -30pt) -- (0pt, -30pt);
    \node[align=center] at (-45pt, -30pt) {initial horizon};
    \draw (0pt, -10pt) -- (0pt, -35pt);
    
    \draw[<-] (0pt, -15pt) -- (10pt, -15pt);
    \draw[->] (40pt, -15pt) -- (50pt, -15pt);
    \node[align=center] at (25pt, -15pt) {\small control \\[-.4em] \small horizon};
    \draw (50pt, -10pt) -- (50pt, -20pt);
    
    \draw[<-] (0pt, -30pt) -- (20pt, -30pt);
    \draw[->] (100pt, -30pt) -- (120pt, -30pt);
    \node[align=center] at (60pt, -30pt) {prediction horizon};
    \draw (120pt, -10pt) -- (120pt, -45pt);
    
    \draw[<-] (-90pt, -40pt) -- (-15pt, -40pt);
    \draw[->] (45pt, -40pt) -- (120pt, -40pt);
    \node[align=center] at (15pt, -40pt) {total horizon};
    
    \draw[->] (-85pt, 0pt) |- (-90pt, 15pt); \node[align=left,anchor=east] at (-90pt, 15pt) {\small step \\[-.3em] \footnotesize $t\!-\!\!L$};
    \draw[->] (-5pt, 0pt) |- (-10pt, 15pt); \node[align=left,anchor=east] at (-10pt, 15pt) {\small step \\[-.3em] \footnotesize $t\!-\!\!1$};
    \draw[->] (5pt, 0pt) |- (10pt, 15pt); \node[align=left,anchor=west] at (10pt, 15pt) {\small step $t$};
    \draw[->] (45pt, 0pt) |- (50pt, 15pt); \node[align=left, anchor=west] at (50pt, 15pt) {\small step \\[-.3em] \footnotesize $t\!+\!\!N_{\rm c}\!\!-\!\!1$};
    \draw[->] (115pt, 0pt) |- (120pt, 15pt); \node[align=left, anchor=west] at (120pt, 15pt) {\small step \\[-.3em] \footnotesize $t\!+\!\!N\!\!-\!\!1$};
    
\end{tikzpicture}

}}
\caption{Time horizons at time $t$.
The initial horizon is a past interval up to time $t-1$. The prediction horizon is a future interval starting at time $t$. 
The total horizon is their union.}
\label{FIG4:Hori}
\end{figure}
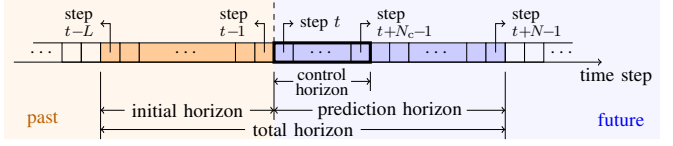

\begin{figure}[t]
\centerline{\resizebox{0.5\textwidth}{!}{

\begin{tikzpicture}
    \tikzstyle{grid} = [draw, rectangle, minimum height=10pt, minimum width=10pt]
    
    \fill[white,draw=gray] (-35pt,40pt) rectangle (-145pt, 25pt);
    \fill[white,draw=gray] (-35pt,40pt) rectangle (190pt, 25pt);
    \fill[white,draw=gray] (-35pt,25pt) rectangle (-145pt, -95pt);
    \fill[white,draw=gray] (-35pt,25pt) rectangle (190pt, -95pt);
    \node[align=left, anchor=west] at (-140pt,32pt) {The column span of:};
    \node[align=left, anchor=west] at (-30pt,33pt) {is the behavior on:};
    
    {
    \node[align=left, anchor=west] at (-140pt,0pt) {$[U_{\rm p}^1; U_{\rm f}^1; Y_{\rm p}^1; Y_{\rm f}^1]$};
    \draw[->] (-25pt,-5pt) -- (150pt, -5pt); \node[] at (160pt, 2pt) {time step};
    \draw[] (0pt, 5pt) -- (-15pt, 5pt); \node[] at (-10pt, 0pt) {$\cdots$};
    \node[grid, fill=red!20] at (5pt, 0pt) {};
    \node[grid, fill=red!20] at (15pt, 0pt) {};
    \node[grid, minimum width=50pt, fill=red!20] at (45pt, 0pt) {}; \node[] at (45pt, 0pt) {$\cdots$};
    \node[grid, fill=red!20] at (75pt, 0pt) {};
    \node[grid] at (85pt, 0pt) {};
    \node[grid, minimum width=20pt] at (100pt, 0pt) {}; \node[] at (100pt, 0pt) {$\cdots$};
    \node[grid] at (115pt, 0pt) {};
    \draw[] (120pt, 5pt) -- (135pt, 5pt); \node[] at (130pt, 0pt) {$\cdots$};
    
    \draw[->] (5pt, 0pt) |- (10pt, 15pt); \node[align=left, anchor=west] at (10pt, 15pt) {\small step \\[-.25em] \small $t^1\!\!-\!\!L$};
    \draw[->] (75pt, 0pt) |- (80pt, 15pt); \node[align=left, anchor=west] at (80pt, 15pt) {\small step \\[-.25em] \small $t^1\!\!+\!\!N\!\!-\!\!1$};
    }
    
    {
    \node[align=left, anchor=west] at (-140pt,-35pt) {$[U_{\rm p}^2; U_{\rm f}^2; Y_{\rm p}^2; Y_{\rm f}^2]$};
    \draw[->] (-25pt,-40pt) -- (150pt, -40pt); \node[] at (160pt, -33pt) {time step};
    \draw[] (0pt, -30pt) -- (-15pt, -30pt); \node[] at (-10pt, -35pt) {$\cdots$};
    \node[grid] at (5pt, -35pt) {};
    \node[grid, fill=red!20] at (15pt, -35pt) {};
    \node[grid, minimum width=50pt, fill=red!20] at (45pt, -35pt) {}; \node[] at (45pt, -35pt) {$\cdots$};
    \node[grid, fill=red!20] at (75pt, -35pt) {};
    \node[grid, fill=red!20] at (85pt, -35pt) {};
    \node[grid, minimum width=20pt] at (100pt, -35pt) {}; \node[] at (100pt, -35pt) {$\cdots$};
    \node[grid] at (115pt, -35pt) {};
    \draw[] (120pt, -30pt) -- (135pt, -30pt); \node[] at (130pt, -35pt) {$\cdots$};
    
    \draw[->] (15pt, -35pt) |- (20pt, -20pt); \node[align=left, anchor=west] at (20pt, -20pt) {\small step \\[-.25em] \small $t^2\!\!-\!\!L$};
    \draw[->] (85pt, -35pt) |- (90pt, -20pt); \node[align=left, anchor=west] at (90pt, -20pt) {\small step \\[-.25em] \small $t^2\!\!+\!\!N\!\!-\!\!1$};
    }
    
    \node[] at (-120pt,-55pt) {\bf $\vdots$};
    \node[] at (-10pt,-55pt) {\bf $\vdots$};
    
    {
    \node[align=left, anchor=west] at (-140pt,-80pt) {$[U_{\rm p}^T; U_{\rm f}^T; Y_{\rm p}^T; Y_{\rm f}^T]$};
    \draw[->] (-25pt,-85pt) -- (150pt, -85pt); \node[] at (160pt, -79pt) {time step};
    \draw[] (0pt, -75pt) -- (-15pt, -75pt); \node[] at (-10pt, -80pt) {$\cdots$};
    \node[grid] at (5pt, -80pt) {};
    \node[grid] at (15pt, -80pt) {};
    \node[grid, minimum width=20pt] at (30pt, -80pt) {}; \node[] at (30pt, -80pt) {$\cdots$};
    \node[grid, fill=red!20] at (45pt, -80pt) {};
    \node[grid, minimum width=60pt, fill=red!20] at (80pt, -80pt) {}; \node[] at (80pt, -80pt) {$\cdots$};
    \node[grid, fill=red!20] at (115pt, -80pt) {};
    \draw[] (120pt, -75pt) -- (135pt, -75pt); \node[] at (130pt, -80pt) {$\cdots$};
    
    \draw[->] (45pt, -80pt) |- (50pt, -65pt); \node[align=left, anchor=west] at (50pt, -62pt) {\small step \\[-.25em] \small $t^T\!\!-\!\!L$};
    \draw[->] (115pt, -80pt) |- (120pt, -65pt); \node[align=left, anchor=west] at (120pt, -62pt) {\small step \\[-.25em] \small $t^T\!\!+\!\!N\!\!-\!\!1$};
    }
    
\end{tikzpicture}

}}
\caption{The column span of the matrix $[U_{\rm p}^\Idx; U_{\rm f}^\Idx; Y_{\rm p}^\Idx; Y_{\rm f}^\Idx]$ is the behavior on the interval $[t^\Idx-L,t^\Idx+N) \cap\integer$.
Note that $\{t^\Idx\}_{\Idx=1}^T$ are consecutive time steps in one period.}
\label{FIG4:MatSpan}
\end{figure}
\begin{figure}[t]
\centerline{\resizebox{0.4\textwidth}{!}{

\begin{tikzpicture}
    \tikzstyle{grid} = [draw, rectangle, minimum height=10pt, minimum width=10pt]
    
    \node[grid] at (-5pt, 0pt) {};
    \draw (-10pt, 5pt) -- (-25pt, 5pt);
    \node at (-20pt, 0pt) {$\cdots$};
    
    \fill [pink!20] (0pt, -20pt) rectangle (70pt, 35pt);
    \node[align=center] at (35pt, 20pt) {initial \\[-.25em] trajectory};
    \draw[dashed] (0pt,-20pt) -- (0pt,35pt);
    \draw[] (0pt,35pt) -- (0pt,45pt);
    \node[grid] at (5pt, 0pt) {};
    \node[grid] at (15pt, 0pt) {};
    \node[grid, minimum width=30pt] at (35pt, 0pt) {};
    \node at (35pt, 0pt) {$\cdots$};
    \node[grid] at (55pt, 0pt) {};
    \node[grid] at (65pt, 0pt) {};
    
    \fill [brown!20] (70pt, -20pt) rectangle (110pt, 35pt);
    \node[align=center] at (90pt, 20pt) {index \\[-.25em] test};
    \draw[dashed] (70pt,-20pt) -- (70pt,35pt);
    \node[grid] at (75pt, 0pt) {};
    \node[grid, minimum width=20pt] at (90pt, 0pt) {}; \node at (90pt, 0pt) {$\cdots$};
    \node[grid] at (105pt, 0pt) {};
    \draw[dashed] (110pt,-20pt) -- (110pt,35pt);
    
    \node[grid, minimum width=30pt] at (125pt, 0pt) {}; \node at (125pt, 0pt) {$\cdots$};
    
    \fill [yellow!15] (140pt, -20pt) rectangle (240pt, 35pt);
    \node[align=center] at (170pt, 20pt) {control \\[-.25em] process};
    \draw[dashed] (140pt,-20pt) -- (140pt,35pt);
    \draw[] (140pt,35pt) -- (140pt,45pt);
    \node[grid] at (145pt, 0pt) {};
    \node[grid] at (155pt, 0pt) {};
    \node[grid] at (165pt, 0pt) {};
    \draw (170pt, 5pt) -- (195pt, 5pt);
    \node at (185pt, 0pt) {$\cdots$};
    
    \draw[<-] (0pt, 40pt) -- (35pt, 40pt);
    \draw[->] (105pt, 40pt) -- (140pt, 40pt);
    \node at (70pt, 40pt) {warm-up process};
    
    \draw[->] (-35pt, -5pt) -- (210pt, -5pt);
    \node[] at (210pt, -12.5pt) {time step};
    
    \draw[->] (5pt, -8pt) -- (5pt, 0pt);
    \node at (10pt, -12pt) {start};
\end{tikzpicture}

}}
\caption{The entire online process. Before the control process is a warm-up process, which includes initial-trajectory collection and an index test.}
\label{FIG4:Online}
\end{figure}
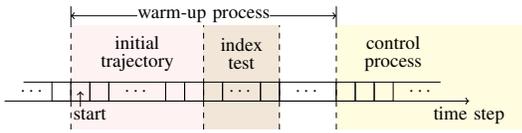

\begin{figure}[t]
\centerline{\resizebox{0.45\textwidth}{!}{

\begin{tikzpicture}
    \tikzstyle{node} = [draw, rectangle, minimum height=20pt, minimum width=30pt, fill = yellow!10]
    
    \node at (-60pt, 75pt) {index $\index=1$};
    \node at (-60pt, 50pt) {index $\index=2$};
    \node at (-60pt, 27.5pt) {$\vdots$};
    \node at (-60pt, 0pt) {index $\index=T$};
    
    \node[node] at (0pt, 75pt) {}; \node at (0pt, 75pt) {$\Delta^1_{t-1}$};
    \node[node] at (0pt, 50pt) {}; \node at (0pt, 50pt) {$\Delta^2_{t-1}$};
    \node at (0pt, 27.5pt) {\Large $\vdots$};
    \node[node, fill=red!20] at (0pt, 0pt) {}; \node at (0pt, 0pt) {$\Delta^T_{t-1}$};
    \node at (0pt, -25pt) {step $t-1$};
    
    \draw[->, red, ultra thick] (15pt, 0pt) -- (50pt, 75pt) -- node[yshift=7pt] {\color{black} $+\delta^1_t$} (85pt, 75pt);
    \draw[->] (15pt, 75pt) -- (50pt, 50pt) -- node[yshift=7pt] {$+\delta^2_t$} (85pt, 50pt);
    \draw[->] (15pt, 50pt) -- (50pt, 25pt) -- (85pt, 25pt);
    \draw[->] (15pt, 25pt) -- (50pt, 0pt) -- node[yshift=7pt] {$+\delta^T_t$} (85pt, 0pt);
    
    \node[node, fill=red!20] at (100pt, 75pt) {}; \node at (100pt, 75pt) {$\Delta^1_t$};
    \node[node] at (100pt, 50pt) {}; \node at (100pt, 50pt) {$\Delta^2_t$};
    \node at (100pt, 27.5pt) {\Large $\vdots$};
    \node[node] at (100pt, 0pt) {}; \node at (100pt, 0pt) {$\Delta^T_t$};
    \node at (100pt, -25pt) {step $t$};
    
    \draw[->] (115pt, 0pt) -- (150pt, 75pt) -- node[yshift=7pt] {$+\delta^1_{t+1}$} (185pt, 75pt);
    \draw[->, red, ultra thick] (115pt, 75pt) -- (150pt, 50pt) -- node[yshift=7pt] {\color{black} $+\delta^2_{t+1}$} (185pt, 50pt);
    \draw[->] (115pt, 50pt) -- (150pt, 25pt) -- (185pt, 25pt);
    \draw[->] (115pt, 25pt) -- (150pt, 0pt) -- node[yshift=7pt] {$+\delta^T_{t+1}$} (185pt, 0pt);
    
    \node[node] at (200pt, 75pt) {}; \node at (200pt, 75pt) {$\Delta^1_{t+1}$};
    \node[node, fill=red!20] at (200pt, 50pt) {}; \node at (200pt, 50pt) {$\Delta^2_{t+1}$};
    \node at (200pt, 27.5pt) {\Large $\vdots$};
    \node[node] at (200pt, 0pt) {}; \node at (200pt, 0pt) {$\Delta^T_{t+1}$};
    \node at (200pt, -25pt) {step $t+1$};
\end{tikzpicture}

}}
\caption{Accumulating the errors $\delta^\Idx_t$ into $\Delta^\Idx_t$.
Each $\Delta^\Idx_t$ is obtained from updating $\Delta^{\prev\Idx}_{t-1}$.
The red boxes and thick arrows show for example the ``correct path'' in which the index $\Idx = \ProperIdx(t)$ is proper at all time $t$.
The ``correct path'' is expected to have the lowest values of both $\delta^\Idx_t$ and $\Delta^\Idx_t$.}
\label{FIG4:Accu}
\end{figure}
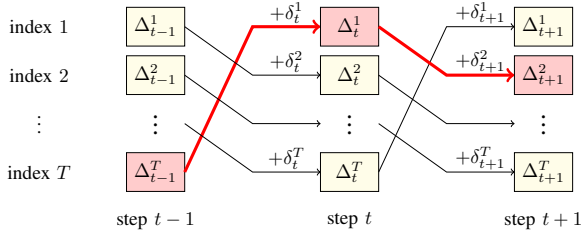

\subsection{Warm-Up Process}
\label{SUBSEC:Algo:WarmUp}

Now we introduce the online process of the algorithm.
Suppose we have collected the offline data in Section \ref{SUBSEC:Algo:OfflineData}.
Before controlling the system, we start by recording an initial trajectory and then finish a so-called ``index test''; see Fig. \ref{FIG4:Online}.
Throughout the warm-up process, no control is occurred and some user-defined input is applied to the system.

\subsubsection{\bf Initial Trajectory}
At the beginning of the online process, we apply no algorithm and simply record an initial trajectory of at least $L$ steps.
This initial trajectory is used to initialize the subsequent procedures.

\subsubsection{\bf Index Test} 
\begin{subequations}
After the initial trajectory is recorded, the next process is an ``index test''.
Although the required proper index $\ProperIdx(t)$ is clearly defined in \eqref{Eq:ProperIdxDef}, this value will be unknown unless $t_{\rm d1}$ is known
during the data collection process, and will generally be
unknown in a practical implementation.
Due to this consideration, a process is required to identify $\ProperIdx(t)$ at current time $t$.
We propose a heuristic index-testing which supports both deterministic and stochastic LTP systems.

At time $t$, consider the recorded past trajectory $w_{[t-L,t)}$ and recall the offline data matrices $U_{\rm p}^\Idx$ and $Y_{\rm p}^\Idx$ in \eqref{Eq:DataMatDef}.
We compute the SVD of the matrix $[U_{\rm p}^\Idx; Y_{\rm p}^\Idx]$ and the error
\begin{align} \label{Eq:PT_Err}
    \delta^\Idx_t := \norm{N_{\rm p}^\Idx {}^\top w_{[t-L,t)}}_2,
\end{align}
where the columns of the matrix $N_{\rm p}^\Idx \in\real^{(m+p)L\times \bullet}$ are the ``non-dominant'' left-singular vectors of $[U_{\rm p}^\Idx; Y_{\rm p}^\Idx]$ that correspond to singular values not exceeding some specified threshold $\sigma_{\rm IT} >0$.
In the deterministic case with $\sigma_{\rm IT} =0$, we have $\delta^\Idx_t = 0$ if, and only if, $w_{[t-L,t)} \in \mathscr{B}^\Sys_{[t^\Idx-L, t^\Idx)}$,\footnote{This is because $\mathscr{B}^\Sys_{[t^\Idx-L,t^\Idx)} = \mathrm{ColSpan} [U_{\rm p}^\Idx; Y_{\rm p}^\Idx] = \mathrm{Null}(N_{\rm p}^\Idx)$, where the first equality is by \eqref{Eq:OfflineData_OnlineBeh} and Lemma \ref{LEMMA:SubBehRepre}, and second equality follows from the definition of $N_{\rm p}^\Idx$ with $\sigma_{\rm IT}=0$.}
so the indices $\Idx$ with $\delta^\Idx_t \neq 0$ are discarded as possibilities of $\ProperIdx(t)$.
In the general case, we regard $\delta^\Idx_t$ as a ``score'' to falsify the hypothesis that the index $\Idx$ is proper at time $t$.

We repeat the above computation on multiple time steps.
Define accumulated errors $\Delta^\Idx_t \in\real$ for $\Idx\in\{1,\ldots,T\}$,
which are initialized to zero and updated in the following way,
\begin{align} \label{Eq:PT_Update}
    \Delta^\Idx_t := \Delta^{\prev\Idx}_{t-1} + \delta^\Idx_t
\end{align}
where $\prev\Idx$ denotes the ``cyclically previous'' index of $\Idx$. \vspace{-.5em}
\begin{align*} \small
    \prev\Idx := \begin{cases}
        \Idx - 1, & \text{for } \Idx \in\{2,\ldots,T\} \\
        T, & \text{for } \Idx =1
    \end{cases}
\end{align*}
\vspace{-1em}

\noindent See Fig. \ref{FIG4:Accu}. The reason we obtain $\Delta^\Idx_t$ by updating $\Delta^{\prev\Idx}_{t-1}$ in \eqref{Eq:PT_Update} is that an index $\Idx$ is proper at time $t$ if, and only if, $\prev\Idx$ is the proper index at the last step $t-1$. Thus, $\Delta^\Idx_t$ is the ``cumulative score'' to falsify the hypothesis that $\Idx=\ProperIdx(t)$, considering all errors computed so far.
The number of iterations $N_{\rm IT}$ of this process is user-specified.
When the algorithm terminates, the index $\Idx$ with the smallest $\Delta^\Idx_t$ is selected as the estimated proper index $\EstProperIdx(t)$ at the current time $t$.
The entire index-testing process is outlined in Algorithm \ref{ALGO:PhaseTest}.

Once we obtain $\EstProperIdx(t)$ at time $t$, the estimated proper index $\EstProperIdx(t')$ for any future time $t'\geq t$ is derived according to
\begin{align*}
    \EstProperIdx(t+1) := \small \begin{cases}
        \EstProperIdx(t) + 1, & \text{if } \EstProperIdx(t) \in\{1,\ldots,T-1\}, \\
        1, & \text{if } \EstProperIdx(t) = T,
    \end{cases} 
\end{align*}
which is the same way that $\ProperIdx(t)$ evolves with time $t$.
\end{subequations}

\begin{algorithm}
\caption{Index Test}
\begin{algorithmic}[1] \label{ALGO:PhaseTest}
    \REQUIRE the time step $t$ and the data matrices $U_{\rm p}^\Idx, Y_{\rm p}^\Idx$ for $\Idx \in \{1, \ldots, T\}$.
    \STATE Compute $N_{\rm p}^\Idx$ as described in Section \ref{SUBSEC:Algo:WarmUp} for all $\Idx$.
    \STATE Initialize the accumulators $\Delta^\Idx_{t-1} = 0$ for all $\Idx$.
    \FOR{$i$ from $1$ to $N_{\rm IT}$}
        \STATE Compute $\delta^\Idx_t$ from \eqref{Eq:PT_Err} for all $\Idx$.
        \STATE Update $\Delta^\Idx_t$ via \eqref{Eq:PT_Update} for all $\Idx$.
        \IF{$i < N_{\rm IT}$} 
        \STATE Input a user-defined $u_t$ to the system $\Sys$.
        \STATE Set $t \gets t+1$.
        \ENDIF
    \ENDFOR
    \ENSURE the estimated proper index $\EstProperIdx(t) \gets \argmin_\Idx \Delta^\Idx_t$.
\end{algorithmic}
\end{algorithm}

\subsection{Online Control Process}
\label{SUBSEC:Algo:Ctrl}

\begin{subequations}
With the proper index identified or known, we can start the control process.
We provide for the LTP system $\Sys$ two alternative controllers, which generalize Data-enabled Predictive Control (DeePC) and Subspace Predictive Control (SPC) methods in the literature.

Let $u^*$ and $y^*$ denote the future input and predicted output respectively.
At step $t$, we consider the quadratic cost
\begin{align} \label{Eq:Cost}
    \textstyle{\sum_{i=t}^{t+N-1}} \norm{y^*_i - r_i}_Q^2 + \norm{u^*_i}_R^2
\end{align}
with cost matrices $Q \succeq 0$ and $R \succ 0$ as parameters,
and constrain the future input-output signal
\begin{align} \label{Eq:I/O_Cons}
    u^*_i \in \mathcal{U}, \quad
    y^*_i \in \mathcal{Y}, \qquad 
    \forall i \in [t,t+N) \cap \integer
\end{align}
with user-defined constraint sets $\mathcal{U} \subseteq \real^m$ and $\mathcal{Y} \subseteq \real^p$.
The \emph{periodic DeePC (P-DeePC) problem} at time $t$ is
\begin{align*} \label{PROB:DeePC_Prob} \tag{P-DeePC} \begin{aligned}[t]
    \mathop{\mathrm{minimize}}_{g, u^*, y^*} \;\; \eqref{Eq:Cost} \;\;
    \mathrm{s.t.} \;\;
    \text{\eqref{Eq:DeePC_Cons} and \eqref{Eq:I/O_Cons}}
\end{aligned} \end{align*}
with an auxiliary variable $g \in \real^h$, where \eqref{Eq:DeePC_Cons} is given as \vspace{-.3em}
\begin{align} \label{Eq:DeePC_Cons}
    {\footnotesize \begin{bmatrix}
        U_{\rm p}^{\EstProperIdx(t)} \\ U_{\rm f}^{\EstProperIdx(t)} \\ Y_{\rm p}^{\EstProperIdx(t)} \\ Y_{\rm f}^{\EstProperIdx(t)}
    \end{bmatrix}} g 
    = \begin{bmatrix}
        u_{[t-L,t)} \\ u^*_{[t,t+N)} \\ y_{[t-L,t)} \\ y^*_{[t,t+N)}
    \end{bmatrix}.
\end{align}
The \emph{periodic SPC (P-SPC) problem} at time $t$
\begin{align} \label{PROB:SPC_Prob} \tag{P-SPC} \begin{aligned}[t]
    \mathop{\mathrm{minimize}}_{u^*, y^*} \;\; \eqref{Eq:Cost} \;\;
    \mathrm{s.t.} \;\;
    \text{\eqref{Eq:SPC_Cons} and \eqref{Eq:I/O_Cons}}
\end{aligned} \end{align}
with \eqref{Eq:SPC_Cons} given as \vspace{-.8em}
\begin{align} \label{Eq:SPC_Cons}
    y^*_{[t,t+N)} = Y_{\rm f}^{\EstProperIdx(t)} 
    \footnotesize \begin{bmatrix}
        U_{\rm p}^{\EstProperIdx(t)} \\ U_{\rm f}^{\EstProperIdx(t)} \\ Y_{\rm p}^{\EstProperIdx(t)}
    \end{bmatrix}^\dagger 
    \normalsize \begin{bmatrix} 
        u_{[t-L,t)} \\ u^*_{[t,t+N)} \\ y_{[t-L,t)}
    \end{bmatrix}.
\end{align}
After solving the optimal future trajectory $w^*_{[t,t+N)}$ from either \eqref{PROB:DeePC_Prob} or \eqref{PROB:SPC_Prob},
we apply the first $N_{\rm c}$ inputs $u^*_{[t, t+N_{\rm c})}$ to the system $\Sys$.
The whole control process is illustrated in Algorithm \ref{ALGO:Ctrl}.

\begin{algorithm}
\caption{Control Process}
\begin{algorithmic}[1] \label{ALGO:Ctrl}
    \REQUIRE the time step $t$, the estimated proper index $\EstProperIdx(t)$, the reference signal $r$ and the data matrices $U_{\rm p}^\Idx, U_{\rm f}^\Idx, Y_{\rm p}^\Idx, Y_{\rm f}^\Idx$ for $\Idx \in \{1, \ldots, T\}$.
    \WHILE{\texttt{true}}
        \STATE Solve $w^*_{[t,t+N)}$ from \eqref{PROB:DeePC_Prob} or \eqref{PROB:SPC_Prob}.
        \STATE Input $u_{[t,t+N_{\rm c})} \gets u^*_{[t,t+N_{\rm c})}$ to the system $\Sys$.
        \STATE Set $t \gets t + N_{\rm c}$ and update $\EstProperIdx(t)$ correspondingly.
    \ENDWHILE
\end{algorithmic}
\end{algorithm}

\subsubsection{\bf Performance Guarantee}
In the deterministic case, both P-DeePC and P-SPC produce the same control actions that one would obtain from traditional MPC applied to the LTP system. The MPC problem for \eqref{Eq:LTV} at time $t$,
\begin{align} \label{PROB:MPC_Prob} \tag{MPC} \begin{aligned}[t]
    \mathop{\mathrm{minimize}}_{x^*, u^*, y^*} \;\; \eqref{Eq:Cost} \;\;
    \mathrm{s.t.} \;\;
    \text{\eqref{Eq:MPC_Cons} and \eqref{Eq:I/O_Cons}}
\end{aligned}\end{align}
where \eqref{Eq:MPC_Cons} is given as follows.
\begin{align} \label{Eq:MPC_Cons} \small \begin{aligned}
    x^*_{i+1} =&\; A_i x^*_i + B_i u^*_i, \quad \forall i \in [t, t+N) \cap \integer
    \\
    y^*_i =&\; C_i x^*_i + D_i u^*_i, \quad \forall i \in [t, t+N) \cap \integer
    \\
    x^*_t =&\; x_t
\end{aligned} \end{align}

\begin{proposition} \label{PROP:Perfm}
    Consider an LTP system $\Sys$ as in \eqref{Eq:LTV} of period $T$. 
    Let $w^{\rm d}_{[t_{\rm d1}, t_{\rm d2}]}$ be offline data from $\Sys$ on interval $[t_{\rm d1}, t_{\rm d2}]$.
    For time step $t \in \integer$ and $L,N \in \natural$, assume that
    \begin{enumerate}[(i)]
        \item $L \geq \lag(\Sys, t-L)$,
        \item $\Sys$ is controllable in the sense of Definition \ref{DEF:Ctrb},
        \item $u^{\rm d}_{[t_{\rm d1}, t_{\rm d2}]}$ is $T$-p.p.e. of order $(\lceil K/T \rceil + \order(\Sys, t_{\rm d1})) T$, with $K := L+N+T-1$, and
        \item $\EstProperIdx(t) = \ProperIdx(t)$.
    \end{enumerate}
    Suppose we know the state $x_t$ and recent trajectory $w_{[t-L,t)}$ of $\Sys$.
    Then,
    \begin{itemize}
        \item the unique optimal trajectory $w^*_{[t,t+N)}$ by \eqref{PROB:DeePC_Prob},
        \item the unique optimal trajectory $w^*_{[t,t+N)}$ by \eqref{PROB:SPC_Prob}, and
        \item the unique optimal trajectory $w^*_{[t,t+N)}$ by \eqref{PROB:MPC_Prob}
    \end{itemize}
    are all same.
\end{proposition}

This result generalizes \cite[Cor. 5.1]{Coulson2019} and \cite[Thm. 1]{Fiedler2021}, which results claim the equivalence of DeePC, SPC and MPC for LTI systems.

\begin{remark}
Our extension of DeePC and SPC to LTP systems is based on the insight that the data collected from an LTP system is equivalent to data collected from an appropriate LTI lifted system. In particular, after stacking LTP-system data into lifted-system data, we can apply the established LTI DDMPC methods and compute control signals for the lifted system, and thereby obtain control signals for the original LTP system. A benefit of our treatment here is that discussion of lifted systems can be entirely omitted once proper behavioral systems concepts are defined directly on the LTP system, as we have done in Section \ref{SEC:Beh}.
\oprocend
\end{remark}

\subsubsection{\bf Regularization}
To adapt our methods for stochastic LTP systems with noisy measurements, we may regularize both P-DeePC and P-SPC.
Regularizing P-DeePC is similar as regularizing DeePC \cite{INTRO:Coulson2019b, INTRO:Coulson2021, INTRO:Huang2021a, INTRO:Huang2021b}. Here we exhibit quadratic regularization, where \eqref{PROB:DeePC_Prob} is modified as follows,
\begin{align*}
    \mathop{\mathrm{minimize}}_{g, u^*, y^*, \sigma_{\rm y}} \;\; \eqref{Eq:Cost} + \lambda_{\rm y} \Vert \sigma_{\rm y} \Vert_2^2 + \lambda_{\rm g} \Vert g \Vert_2^2 \;\; 
    \mathrm{s.t.} \;\;
    \text{\eqref{Eq:RegDeePC_Cons} and \eqref{Eq:I/O_Cons}}
\end{align*}
with a slack variable $\sigma_{\rm y} \in \real^{pL}$, positive parameters $\lambda_{\rm y}, \lambda_{\rm g}$, and \eqref{Eq:RegDeePC_Cons} a modified constraint from \eqref{Eq:DeePC_Cons}.
\begin{align} \label{Eq:RegDeePC_Cons}
    {\footnotesize \begin{bmatrix}
        U_{\rm p}^{\EstProperIdx(t)} \\ U_{\rm f}^{\EstProperIdx(t)} \\ Y_{\rm p}^{\EstProperIdx(t)} \\ Y_{\rm f}^{\EstProperIdx(t)}
    \end{bmatrix}} g 
    = \begin{bmatrix}
        u_{[t-L,t)} \\ u^*_{[t,t+N)} \\ y_{[t-L,t)} \\ y^*_{[t,t+N)}
    \end{bmatrix} + \begin{bmatrix}
        0 \\ 0 \\ \sigma_{\rm y} \\ 0
    \end{bmatrix}.
\end{align}
To regularize P-SPC, in the computation of the pseudo-inverse in \eqref{Eq:SPC_Cons}, we treat as zero the singular values smaller than a selected threshold $\sigma_{\rm SPC}$; the remainder of the settings in regularized P-SPC are same as in P-SPC.
\end{subequations}
\section{Simulations} \label{SEC:Simu}

We illustrate the algorithm proposed in Section \ref{SEC:Alg} and its robustness to noisy data via numerical example. Consider the mass-spring-damper system in Fig. \ref{FIG5:Model}. 

\tikzstyle{spring} = [thick, decorate, decoration = {coil,
    segment length = 1.5mm,
    amplitude = 1.2mm,
    aspect = 0.5,
    post length = 3mm,
    pre length = 3mm}]
\tikzstyle{damper} = [thick,
    decoration = {markings, 
    mark connection node = dmp, 
    mark = at position 0.5 with {
        \node (dmp) [ 
        inner sep = 0pt, 
        transform shape, 
        rotate = -90, 
        minimum width = 7pt, 
        minimum height = 8pt, 
        draw = none] {};
        \draw [thick] ($(dmp.north east) + (4pt, 0pt)$)
        -- (dmp.south east) 
        -- (dmp.south west) 
        -- ($(dmp.north west) + (4pt, 0pt)$);
        \draw [ultra thick] ($(dmp.north) + (0, -3pt)$) 
        -- ($(dmp.north) + (0, 3pt)$);}
    }, decorate]

\begin{figure}[ht!]
\centerline{\resizebox{0.4\textwidth}{!}{

\begin{tikzpicture}
    \tikzstyle{mass} = [draw, rectangle, thick, minimum height=40pt, minimum width=30pt, fill=orange!30]
    
    \draw[thick] (0pt, 0pt) -- (0pt, 115pt);
    \fill[pattern = north west lines] (0pt, 0pt) rectangle (-5pt, 115pt);
    \draw[thick] (270pt, 0pt) -- (270pt, 115pt);
    \fill[pattern = north west lines] (270pt, 0pt) rectangle (275pt, 115pt);
    
    \node[mass, minimum height=100pt] at (80pt, 50pt) {$m_1$};
    \draw[dashed] (80pt, 100pt) -- (80pt, 115pt);
    \draw[->] (80pt, 105pt) -- node [pos=1, yshift=5pt] {$x_1$} (100pt, 105pt);
    \draw[->] (45pt, 50pt) -- node [pos=-.3] {$F$} (64pt, 50pt);
    
    \node[mass] at (170pt, 20pt) {$m_3$};
    \draw[dashed] (170pt, 40pt) -- (170pt, 55pt);
    \draw[->] (170pt, 45pt) -- node [pos=1, yshift=5pt] {$x_3$} (190pt, 45pt);
    
    \node[mass] at (170pt, 80pt) {$m_2$};
    \draw[dashed] (170pt, 100pt) -- (170pt, 115pt);
    \draw[->] (170pt, 105pt) -- node [pos=1, yshift=5pt] {$x_2$} (190pt, 105pt);
    
    \draw[spring] (0pt, 70pt) -- node [yshift=9pt] {$k_1$} (65pt, 70pt);
    \draw[damper] (0pt, 30pt) -- node [yshift=-9pt] {$c_1$} (65pt, 30pt);
    
    \draw[spring] (95pt, 30pt) -- node [yshift=9pt] {$k_3$} (155pt, 30pt);
    \draw[damper] (95pt, 10pt) -- node [yshift=-9pt] {$c_3$} (155pt, 10pt);
    
    \draw[spring] (95pt, 90pt) -- node [yshift=9pt] {$k_2$} (155pt, 90pt);
    \draw[damper] (95pt, 70pt) -- node [yshift=-9pt] {$c_2$} (155pt, 70pt);
    
    \draw[spring] (185pt, 30pt) -- node [yshift=9pt] {$k_5$} (240pt, 30pt);
    \draw[damper] (185pt, 10pt) -- node [yshift=-9pt] {$c_5$} (270pt, 10pt);
    \draw[dashed] (240pt, 30pt) -- (240pt, 55pt);
    \draw[->] (240pt, 45pt) -- node [pos=1, yshift=5pt] {$x_5$} (260pt, 45pt);
    
    \draw[spring] (185pt, 90pt) -- node [yshift=9pt] {$k_4$} (240pt, 90pt);
    \draw[damper] (185pt, 70pt) -- node [yshift=-9pt] {$c_4$} (270pt, 70pt);
    \draw[dashed] (240pt, 90pt) -- (240pt, 115pt);
    \draw[->] (240pt, 105pt) -- node [pos=1, yshift=5pt] {$x_4$} (260pt, 105pt);
\end{tikzpicture}

}}
\caption{A spring-mass-damper model for simulation.}
\label{FIG5:Model}
\end{figure}
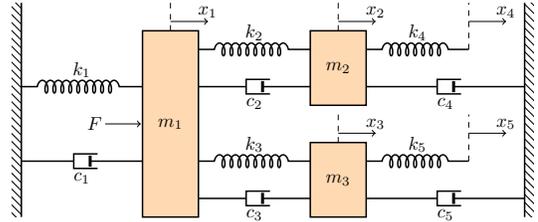

The control objective is reference tracking for the positions $(x_1,x_2,x_3)$ of the three masses. There are three control inputs: the force $F$ applied to the mass $m_1$, and the end positions $x_4$ and $x_5$ of the free ends of the springs $k_4$ and $k_5$. 
The stiffness and damping parameters $k_i$ and $c_i$ are periodic functions of time, given in Table \ref{TABLE1:Phys}, and each has a period of 1 second. We discretize the system with a sampling time $0.2 \rm s$, and thus the period of the discretized system is $T = 5$. A process noise $w_t \overset{\text{i.i.d}}{\sim} \mathcal{N}(0_{6\times 1}, \sigma^2 I_6)$ and a measurement noise $v_t \overset{\text{i.i.d}}{\sim} \mathcal{N}(0_{3\times 1}, \sigma^2 I_3)$ are added to the discrete-time model, with noise amplitude $\sigma^2 = 10^{-3}$.
The control and index-test parameters are selected in Table \ref{TABLE2:CtrlParam}.

For collection of offline data, we apply a random input signal $u^{\rm d}_t \overset{\text{i.i.d.}}{\sim} \mathcal{N}(0_{3\times 1}, I_3)$ and measure the resulting positions $(x_1,x_2,x_3)$. The online process starts at time $t=0$; in the warm-up process, the input is random $u_t \overset{\text{i.i.d.}}{\sim} \mathcal{N}(0_{3\times 1}, I_3/10)$, so that heuristically the index test gives a correct result.
After recording an initial trajectory on interval $[0,29]$, we start the index test at time $t=29$, and terminate the process at time $t=40$ (as $N_{\rm IT}=12$).
In our simulation, the proper index $\ProperIdx(t)$ was identified correctly.

\begin{figure}[ht!]
\centerline{\resizebox{0.5\textwidth}{!}{\includegraphics{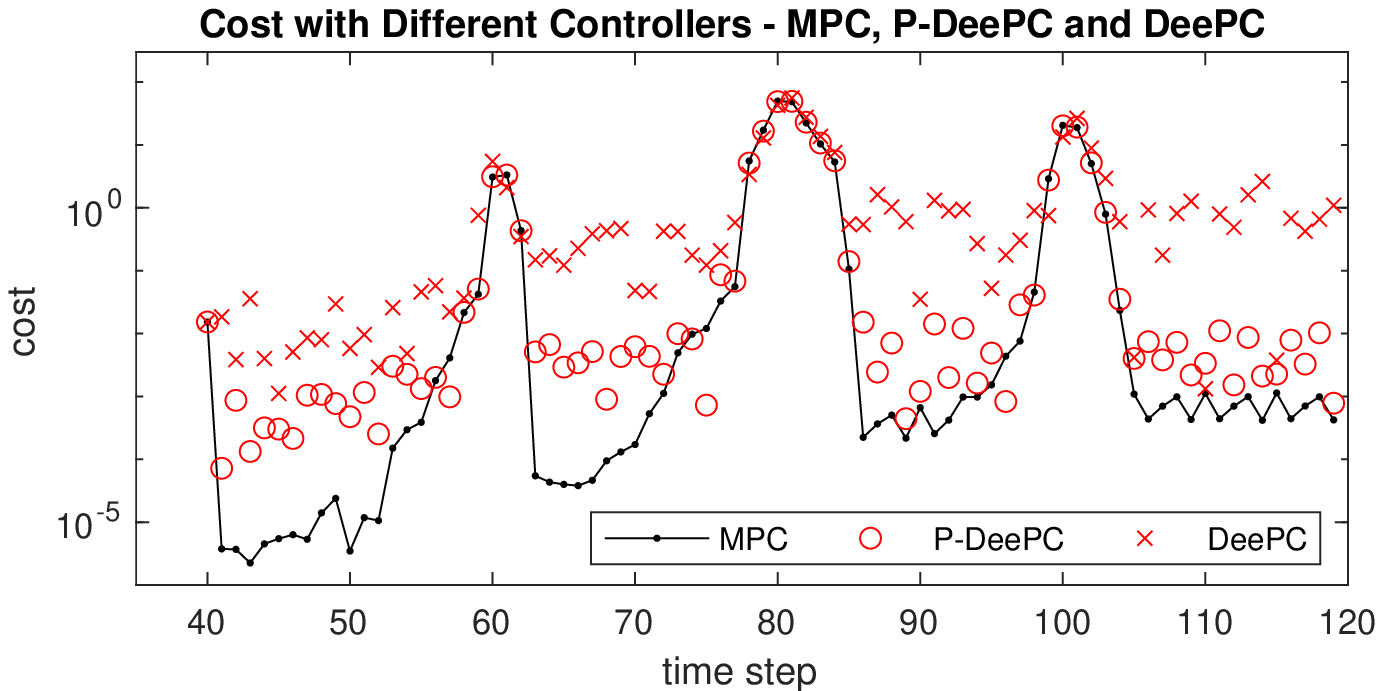}}}
\centerline{\resizebox{0.5\textwidth}{!}{\includegraphics{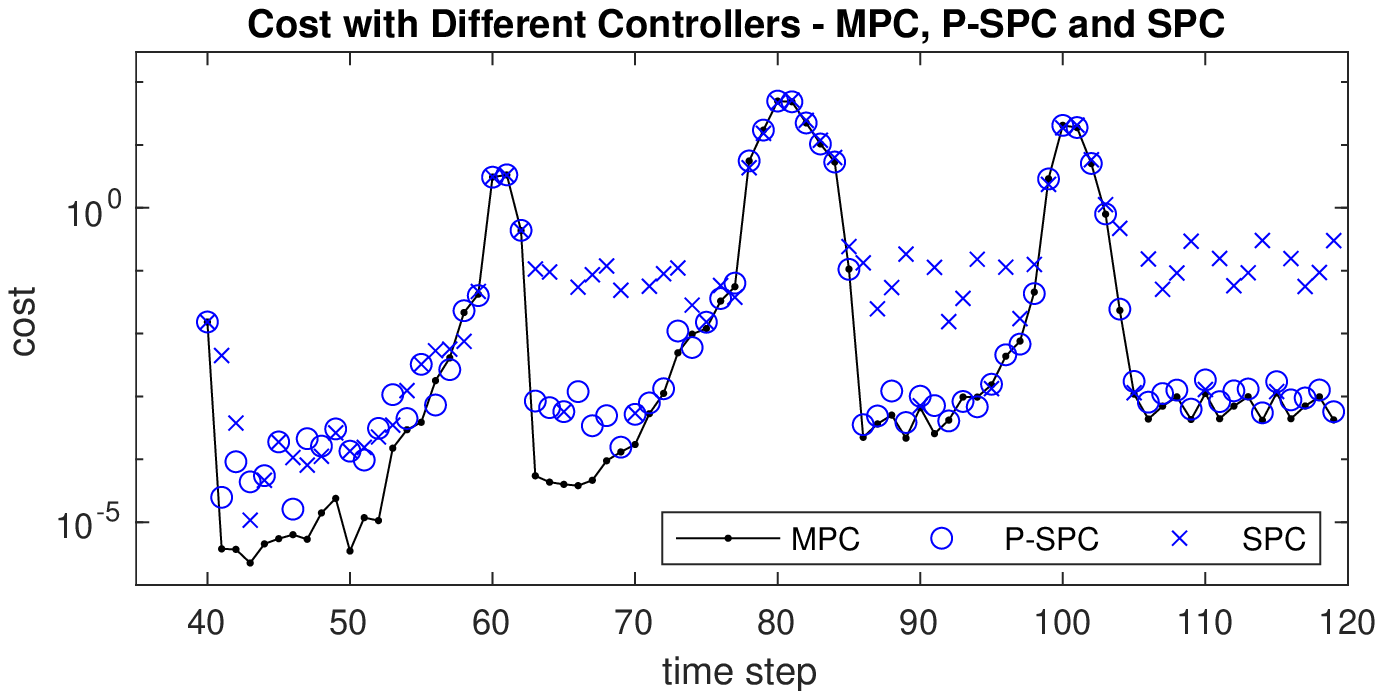}}}
\caption{The cost of the simulated LTP system, using MPC, regularized P-DeePC, regularized P-SPC, regularized DeePC of LTI systems and regularized SPC of LTI systems, respectively.}
\label{FIG5:Plot}
\end{figure}

We start control at time $t = 40$, and apply sequential changes in the reference signals given by $r_t = [0;0;0]$ for $40 \leq t < 60$, $r_t = [5;0;0]$ for $60 \leq t < 80$, $r_t = [5;15;0]$ for $80 \leq t < 100$, and $r_t = [5;15;-10]$ for $t \geq 100$. We evaluate the control performance via the one-step cost $\Vert y_t - r_t \Vert_Q^2 + \Vert u_t \Vert_R^2$, and the results are shown in Fig. \ref{FIG5:Plot}.

For comparison purposes, we also plot the closed-loop responses under (i) MPC using a perfect system model with full-state measurements, and (ii) the regularized DeePC and regularized SPC methods for LTI systems. For the latter, the settings are the same as for P-DeePC (resp. P-SPC), except that we replace the matrices {\small $U_{\rm p}^{\EstProperIdx(t)}, U_{\rm f}^{\EstProperIdx(t)}, Y_{\rm p}^{\EstProperIdx(t)}, Y_{\rm f}^{\EstProperIdx(t)}$} in \eqref{Eq:RegDeePC_Cons} (resp. \eqref{Eq:SPC_Cons}) by $U_{\rm p}^1, U_{\rm f}^1, Y_{\rm p}^1, Y_{\rm f}^1$ respectively, i.e., we use a single set of data matrices at all time $t$. Around the step changes of the reference signal, all controllers have comparable performances with similar cost values.
For the steady-state performance when the reference signal stays constant, the proposed regularized P-DeePC (resp. P-SPC) method outperforms the direct use of regularized DeePC (resp. SPC) of LTI systems.
This significant difference indicates the necessity of using different sets of data matrices $U_{\rm p}^\Idx, U_{\rm f}^\Idx, Y_{\rm p}^\Idx, Y_{\rm f}^\Idx$ with different indices $\Idx$ as in \eqref{Eq:RegDeePC_Cons} and \eqref{Eq:SPC_Cons} for P-DeePC and P-SPC respectively at different time steps.

\section{Conclusion and Future Work} \label{SEC:Con}

We proposed a DDMPC algorithm for unknown LTP systems with known periods. For deterministic LTP systems, the method is equivalent to classical MPC, but without the requirement of a parametric model. The approach is supported by extensions of results from behavioral systems theory to LTV and LTP systems. Simulation results provide evidence that the approach is robust to measurement noise and stochasticity, and that it significantly outperforms a naive application of data-driven LTI control methods. 

There are several open directions for future work. First, as our design requires a priori knowledge of the period $T$, relaxing this assumption is of interest, as is investigating the robustness of the approach to errors in the selected period. 
Second, we will seek to develop a rigorous performance guarantee for the ``index test'' outlined in Section \ref{SUBSEC:Algo:WarmUp}. 
Finally, we note that there remain open questions in the behavioral theory of LTP systems, such as what relationships can be established between the behaviors of the $T$ different lifted systems arising from a given LTP system.

\afterpage{

\begin{table}[t] 
\caption{Physical Quantities of the Simulated System}
\centering 
\begin{tabular}{@{}lll@{}} 
\toprule
    \textbf{Quantity (Unit)} & \textbf{Symbol} & \textbf{Value$^{\rm a}$} 
    \\ \midrule
    Mass ($\rm kg$) & $m_1, m_2, m_3$ & $6, 4, 3$ respectively \\
    \cmidrule{1-1}
    \multirow{3}{*}{\begin{tabular}{@{}l@{}}
        Spring Stretch \\ ($\rm N / m$)
    \end{tabular}} 
    & $k_1$ & $10 - 4\sin(2 \pi t) + 2\sin(4 \pi t)$ \\
    & $k_2, k_3$ & $7 - 3\cos(4 \pi t)$ \\
    & $k_4, k_5$ & $4 - 2\sin(4 \pi t)$ \\ 
    \cmidrule{1-1}
    \multirow{3}{*}{\begin{tabular}{@{}l@{}}
        Viscosity \\ ($\rm N \cdot s / m$) 
    \end{tabular}}
    & $c_1$ & $9 + 3\sin(2 \pi t)$ \\
    & $c_2, c_3$ & $5 + 2\cos(2 \pi t)$ \\
    & $c_4, c_5$ & $15$ \\ 
    \bottomrule
    \multicolumn{3}{@{}l@{}}{$^{\rm a}$Variable $t$ is the time in seconds.}
\end{tabular}
\label{TABLE1:Phys}
\end{table}

\begin{table}[t] 
\caption{Parameters for Control and Index Test}

\begin{center} \begin{tabular}{@{}ll@{}} 
\toprule
    \textbf{Parameter} & \textbf{Value} 
    \\ 
    \midrule
    initial horizon length $L$ & $30$ \\
    prediction horizon length $N$ & $30$ \\
    control horizon length $N_{\rm c}$ & $1$ \\ \midrule
    tracking cost matrix $Q$ in \eqref{Eq:Cost} & $\mathrm{diag}(1, 1, 1)$ \\
    input cost matrix $R$ in \eqref{Eq:Cost} & $\mathrm{diag}(10^{-6}, 10^{-4}, 10^{-4})$ \\ \midrule
    input constraint set $\mathcal{U}$ in \eqref{Eq:I/O_Cons} & $[-8,8]\times[-3,3]\times[-3,3]$ \\
    output constraint set $\mathcal{Y}$ in \eqref{Eq:I/O_Cons} & $[-20,20]^3$ \\ \midrule
    iteration number $N_{\rm IT}$ in the index test & $12$ \\
    SV threshold $\sigma_{\rm IT}$ in the index test & $1$ \\
    \midrule
    regularization parameter $\lambda_{\rm y}$ for P-DeePC & $10^6$ \\
    regularization parameter $\lambda_{\rm g}$ for P-DeePC & $10^{-3}$ \\
    SV threshold $\sigma_{\rm SPC}$ in regularized P-SPC & $0.5$ \\
    \bottomrule 
\end{tabular} \end{center}

\label{TABLE2:CtrlParam}
\end{table}

}

\newpage
\section*{Appendix: Proofs}

\begin{pfof}{Lemma \ref{LEMMA:BehColSpan}}
    The definition of $\mathscr{B}^\Sys_{[t_1,t_2]}$ can be rewritten as
    \begin{align*}
        \mathscr{B}^\Sys_{[t_1,t_2]}
        = \left\{ 
        {\footnotesize \begin{bmatrix}
            u_{[t_1,t_2]} \\ y_{[t_1,t_2]}
        \end{bmatrix}}\,\,
        \middle|\,\,
            \exists x_{t_1} \;\text{s.t.}\;\eqref{Eq:SysMatNotation:OutputSolution}\;\text{holds}
        \right\}.
    \end{align*}
    The result now follows immediately by eliminating $y_{[t_1,t_2]}$ above using \eqref{Eq:SysMatNotation:OutputSolution}.
\end{pfof}

\begin{pfof}{Corollary \ref{COROLLARY:BehDim}}
    The result follows from Lemma \ref{LEMMA:BehColSpan}, with $\begin{bsmallmatrix} 0 & I \\ \mathscr{O}^{t_2}_{t_1} & \mathscr{I}^{t_2}_{t_1} \end{bsmallmatrix}$ block-triangular.
\end{pfof}

\begin{pfof}{Lemma \ref{LEMMA:SubBehRepre}}
    In this proof, let
    \begin{align*}
        \mathcal{C}^{t_1}_{t_0} := \mathrm{ColSpan} [U_{t_0}; \ldots; U_{t_1}; Y_{t_0}; \ldots; Y_{t_1}].
    \end{align*}
    Construct a truncation mapping $f_{\rm trc} : \real^{(t_2-t_0+1)(m+p)} \to \real^{(t_1-t_0+1)(m+p)}$
    \begin{align*} \footnotesize
        f_{\rm trc} \left( \begin{bmatrix} u_{[t_0, t_2]} \\ y_{[t_0, t_2]} \end{bmatrix} \right) := \begin{bmatrix} u_{[t_0, t_1]} \\ y_{[t_0, t_1]} \end{bmatrix}
    \end{align*}
    for all input-output signals $(u, y)$.
    Through this mapping, $\mathcal{C}^{t_1}_{t_0}$ is the image of $\mathcal{C}^{t_2}_{t_0}$, and the image of $\mathscr{B}^\Sys_{[t_0, t_2]}$ is
    \begin{align*} \footnotesize
        \mathcal{B}_{t_0}^{t_1} :=
        \left\{ \begin{bmatrix} u_{[t_0, t_1]} \\ y_{[t_0, t_1]} \end{bmatrix} 
        \middle|
        \begin{bmatrix} u_{[t_0, t_2]} \\ y_{[t_0, t_2]} \end{bmatrix} \in \mathscr{B}^\Sys_{[t_0,t_2]} 
        \right\}.
    \end{align*}
    Now we show that $\mathcal{B}_{t_0}^{t_1} = \mathscr{B}^\Sys_{[t_0, t_1]}$.
    From Lemma \ref{LEMMA:BehColSpan}, we have
    \begin{align*} \footnotesize
        \mathscr{B}^\Sys_{[t_0,t_2]} 
        =&\; \mathrm{ColSpan} \left[ \begin{array}{c|c}
            0 & I \\ \hline
            \mathscr{O}_{t_0}^{t_2} & \mathscr{I}_{t_0}^{t_2}
        \end{array} \right]
        \\=&\; \mathrm{ColSpan} \left[ \begin{array}{c|cc} 
            0 & I & 0 \\
            0 & 0 & I \\ \hline
            \mathscr{O}_{t_0}^{t_1} & \mathscr{I}_{t_0}^{t_1} & 0 \\
            \mathscr{O}_{t_1+1}^{t_2} \Phi_{t_0}^{t_1+1} & \mathscr{O}_{t_1+1}^{t_2} \mathscr{C}_{t_0}^{t_1} & \mathscr{I}_{t_1+1}^{t_2}
        \end{array} \right]
    \end{align*}
    where the second equality can be verified by expanding $\Phi^{t_2}_{t_1}, \mathscr{C}^{t_2}_{t_1}, \mathscr{O}^{t_2}_{t_1}, \mathscr{I}^{t_2}_{t_1}$ into system matrices via \eqref{Eq:SysMatNotation_Phi}, \eqref{Eq:SysMatNotation_C}, \eqref{Eq:SysMatNotation_O}, \eqref{Eq:SysMatNotation_I} respectively.
    Through the truncating operation $f_{\rm trc}$, the image $\mathcal{B}_{t_0}^{t_1}$ of $\mathscr{B}^\Sys_{[t_0,t_2]}$ is therefore
    \begin{align*}
        \mathcal{B}_{t_0}^{t_1} = \mathrm{ColSpan} \begin{bmatrix}
            0 & I \\
            \mathscr{O}_{t_0}^{t_1} & \mathscr{I}_{t_0}^{t_1}
        \end{bmatrix}
    \end{align*}
    which equals $\mathscr{B}^\Sys_{[t_0, t_1]}$ by Lemma \ref{LEMMA:BehColSpan}, so $\mathscr{B}^\Sys_{[t_0, t_1]}$ is the image of $\mathscr{B}^\Sys_{[t_0, t_2]}$ by operation $f_{\rm trc}$.
    The result $\mathscr{B}^\Sys_{[t_0, t_1]} = \mathcal{C}_{t_0}^{t_1}$ follows because the images of equal sets $\mathscr{B}^\Sys_{[t_0, t_2]} = \mathcal{C}_{t_0}^{t_2}$ are equal.
\end{pfof}

\begin{pfof}{Lemma \ref{LEMMA:DiffModelSameBeh}}
    \textbf{If.} From \eqref{Eq:LEMMA:DiffModelSameBeh:Cond2}, with $\mathcal{O}$ spanned by $\mathscr{O}^{t_2}_{t_1}$, there exists a matrix $\mathcal{L} \in\real^{n\times n}$ such that $\mathscr{I}^{t_2}_{t_1} - \mathscr{\bar I}^{t_2}_{t_1} = \mathscr{O}^{t_2}_{t_1} \mathcal{L}$. Hence,
    \begin{align*}
        &\; \small
        \mathrm{ColSpan} \begin{bmatrix} 
            0 & I \\ 
            \mathscr{O}^{t_2}_{t_1} & \mathscr{I}^{t_2}_{t_1}
        \end{bmatrix}
        = \mathrm{ColSpan} \begin{bmatrix} 
            0 & I \\ 
            \mathscr{O}^{t_2}_{t_1} & \mathscr{\bar I}^{t_2}_{t_1} + \mathscr{O}^{t_2}_{t_1} \mathcal{L}
        \end{bmatrix}
        \\&\; \small
        \overset{\substack{\text{by column}\\ \text{operation}}}{=\joinrel=} \mathrm{ColSpan}
        \begin{bmatrix} 
            0 & I \\ 
            \mathscr{O}^{t_2}_{t_1} & \mathscr{\bar I}^{t_2}_{t_1}
        \end{bmatrix}
        \overset{\text{via \eqref{Eq:LEMMA:DiffModelSameBeh:Cond1}}}{=\joinrel=} \mathrm{ColSpan} \begin{bmatrix}
            0 & I \\ 
            \mathscr{\bar O}^{t_2}_{t_1} & \mathscr{\bar I}^{t_2}_{t_1}
        \end{bmatrix}
    \end{align*}
    which implies $\mathscr{B}^\Sys_{[t_1,t_2]} = \mathscr{B}^{\bar \Sys}_{[t_1,t_2]}$ via Lemma \ref{LEMMA:BehColSpan}.
    \textbf{Only if.}
    By Lemma \ref{LEMMA:BehColSpan}, the matrices
    \begin{align*}
        \mathcal{V} := {\small \begin{bmatrix} 0 & I \\ \mathscr{O}^{t_2}_{t_1} & \mathscr{I}^{t_2}_{t_1} \end{bmatrix}}
        \quad \text{and} \quad
        \mathcal{\bar V} := {\small \begin{bmatrix} 0 & I \\ \mathscr{\bar O}^{t_2}_{t_1} & \mathscr{\bar I}^{t_2}_{t_1} \end{bmatrix}}
    \end{align*}
    have the same column span.
    As $\mathrm{ColSpan}(\mathcal{V}) \supseteq \mathrm{ColSpan}(\mathcal{\bar V})$, there exist matrices $\mathcal{M} \in\real^{n\times n}$, $\mathcal{N} \in\real^{n\times mL}$, $\mathcal{P} \in\real^{mL\times n}$ and $\mathcal{Q} \in\real^{mL \times mL}$, where $L := t_2-t_1+1$, such that
    \begin{align*} \small
        \begin{bmatrix} 
            0 & I \\ 
            \mathscr{O}^{t_2}_{t_1} & \mathscr{I}^{t_2}_{t_1} 
        \end{bmatrix}
        \begin{bmatrix}
            \mathcal{M} & \mathcal{N} \\
            \mathcal{P} & \mathcal{Q}
        \end{bmatrix}
        = \begin{bmatrix} 
            0 & I \\ 
            \mathscr{\bar O}^{t_2}_{t_1} & \mathscr{\bar I}^{t_2}_{t_1} 
        \end{bmatrix}.
    \end{align*}
    Compute the left-hand side above and compare the result to the right-hand side, and then we have $\mathcal{P} = 0$, $\mathcal{Q} = I$ and 
    \begin{align*}
        \mathscr{O}^{t_2}_{t_1} \mathcal{M} = \mathscr{\bar O}^{t_2}_{t_1}, \qquad
        \mathscr{I}^{t_2}_{t_1} - \mathscr{\bar I}^{t_2}_{t_1} = - \mathscr{O}^{t_2}_{t_1} \mathcal{N}.
    \end{align*}
    Recall $\mathcal{O} = \mathrm{ColSpan}(\mathscr{O}^{t_2}_{t_1})$ and let $\mathcal{\bar O} := \mathrm{ColSpan}(\mathscr{\bar O}^{t_2}_{t_1})$.
    The above equations imply that $\mathrm{ColSpan}(\mathscr{I}^{t_2}_{t_1} - \mathscr{\bar I}^{t_2}_{t_1})$ is a subspace of $\mathcal{O}$, i.e. we have \eqref{Eq:LEMMA:DiffModelSameBeh:Cond2},
    and $\mathcal{\bar O}$ is a subspace of $\mathcal{O}$.
    Similarly, as $\mathrm{ColSpan}(\mathcal{V}) \subseteq \mathrm{ColSpan}(\mathcal{\bar V})$,
    we have the converse result that $\mathcal{O}$ is a subspace of $\mathcal{\bar O}$, so $\mathcal{O} = \mathcal{\bar O}$, i.e. \eqref{Eq:LEMMA:DiffModelSameBeh:Cond1} is obtained.
\end{pfof}

\begin{pfof}{Lemma \ref{LEMMA:LagIniCond}}
    \begin{subequations}
    With abuse of notation, we let
    \begin{align*}
        \Phi_{\rm p} :=&\; \Phi^{t}_{t-L}, &
        \mathscr{C}_{\rm p} :=&\; \mathscr{C}^{t-1}_{t-L},
        \\
        \mathscr{O}_{\rm p} :=&\; \mathscr{O}^{t-1}_{t-L}, &
        \mathscr{O}_{\rm f} :=&\; \mathscr{O}^{t+N-1}_{t}, &
        \mathscr{O}_{\rm pf} :=&\; \mathscr{O}^{t+N-1}_{t-L},
        \\
        \mathscr{I}_{\rm p} :=&\; \mathscr{I}^{t-1}_{t-L}, &
        \mathscr{I}_{\rm f} :=&\; \mathscr{I}^{t+N-1}_{t}, &
        \mathscr{I}_{\rm pf} :=&\; \mathscr{I}^{t+N-1}_{t-L},
    \end{align*}
    in this proof, where the subscript ``p'' stands for the past interval $[t-L,t) \cap \integer$, ``f'' for the future interval $[t,t+N) \cap \integer$, and ``pf'' their union.
    One can verify that
    \begin{align} \label{Eq:PROOF:LagIniCond:0}
        \mathscr{O}_{\rm pf} = \begin{bmatrix}
            \mathscr{O}_{\rm p} \\ \mathscr{O}_{\rm f} \Phi_{\rm p}
        \end{bmatrix}, \qquad
        \mathscr{I}_{\rm pf} = \begin{bmatrix}
            \mathscr{I}_{\rm p} & 0 \\
            \mathscr{O}_{\rm f} \mathscr{C}_{\rm p} & \mathscr{I}_{\rm f}
        \end{bmatrix},
    \end{align}
    by expanding $\Phi^{t_2}_{t_1}, \mathscr{C}^{t_2}_{t_1}, \mathscr{O}^{t_2}_{t_1}, \mathscr{I}^{t_2}_{t_1}$ into system matrices via \eqref{Eq:SysMatNotation_Phi}, \eqref{Eq:SysMatNotation_C}, \eqref{Eq:SysMatNotation_O}, \eqref{Eq:SysMatNotation_I} respectively.
    
    (i): Since $w_{[t-L,t)} \in \mathscr{B}^\Sys_{[t-L, t)}$, by definition there exists some initial state $x_{t-L}$ such that $y_{[t-L,t)}$ is the output resulting from input $u_{[t-L,t)}$, with the resulting state $x_t$ at time $t$. From \eqref{Eq:SysMatNotation:StateSolution} and \eqref{Eq:SysMatNotation:OutputSolution} we have
    \begin{align}
        \label{Eq:PROOF:LagIniCond:1-1}
        x_t =&\; \Phi_{\rm p} x_{t-L} + \mathscr{C}_{\rm p} u_{[t-L,t)},
        \\
        \label{Eq:PROOF:LagIniCond:1-2}
        y_{[t-L,t)} =&\; \mathscr{O}_{\rm p} x_{t-L} + \mathscr{I}_{\rm p} u_{[t-L,t)}.
    \end{align}
    Thus, via \eqref{Eq:SysMatNotation:OutputSolution} the resulting output with the initial state $x_t$ and input $u^*_{[t,t+N)}$ is
    \begin{align} \label{Eq:PROOF:LagIniCond:1-3}
        y^*_{[t,t+N)} = \mathscr{O}_{\rm f} x_t + \mathscr{I}_{\rm f} u^*_{[t,t+N)},
    \end{align}
    which is an existing $y^*_{[t,t+N)}$ that satisfies \eqref{Eq:LEMMA:LagIniCond:Beh}.
    
    (ii): Substituting \eqref{Eq:PROOF:LagIniCond:1-1} into \eqref{Eq:PROOF:LagIniCond:1-3}, we have
    \begin{align} \label{Eq:PROOF:LagIniCond:2-1}
        y^*_{[t,t+N)} =&\; \mathscr{O}_{\rm f} (\Phi_{\rm p} x_{t-L}  + \mathscr{C}_{\rm p} u_{[t-L,t)}) + \mathscr{I}_{\rm f} u^*_{[t,t+N)}.
    \end{align}
    To show the uniqueness of $y^*_{[t,t+N)}$, it suffices to show that the term $\mathscr{O}_{\rm f} \Phi_{\rm p} x_{t-L}$ in \eqref{Eq:PROOF:LagIniCond:2-1} is unique, although $x_{t-L}$ may not be unique.
    Since $L \geq \lag(\Sys, t-L)$, it follows from the definition of lag in \eqref{Eq:OrderLag_LTV} and the definitions of $\mathscr{O}_{\rm p}, \mathscr{O}_{\rm pf}$ that
    \begin{align*}
        \rank(\mathscr{O}_{\rm p})
        = \rank(\mathscr{O}_{\rm pf})
    \end{align*}
    Due to the rank equality above and the segmentation $\mathscr{O}_{\rm pf} = [\mathscr{O}_{\rm p}; \mathscr{O}_{\rm f} \Phi_{\rm p}]$ in \eqref{Eq:PROOF:LagIniCond:0}, we conclude that the rows in $\mathscr{O}_{\rm f} \Phi_{\rm p}$ are linearly dependent to the rows in $\mathscr{O}_{\rm p}$,
    so there exists a matrix $\mathcal{M}\in \real^{pN \times pL}$ such that
    \begin{align} 
        \label{Eq:PROOF:LagIniCond:2-2}
        \mathscr{O}_{\rm f} \Phi_{\rm p} = \mathcal{M} \mathscr{O}_{\rm p}.
    \end{align}
    Then, we obtain that
    \begin{align*}
        \mathscr{O}_{\rm f} \Phi_{\rm p} x_{t-L}
        \overset{\text{via \eqref{Eq:PROOF:LagIniCond:2-2}}}{=\joinrel=} &\;
        \mathcal{M} \mathscr{O}_{\rm p} x_{t-L}
        = \mathcal{M} \mathscr{O}_{\rm p} \mathscr{O}_{\rm p}{}^\dagger \mathscr{O}_{\rm p} x_{t-L}
        \\ \overset{\text{via \eqref{Eq:PROOF:LagIniCond:2-2}}}{=\joinrel=}
        &\;
        \mathscr{O}_{\rm f} \Phi_{\rm p} \mathscr{O}_{\rm p}{}^\dagger \mathscr{O}_{\rm p} x_{t-L}
        \\ \overset{\text{via \eqref{Eq:PROOF:LagIniCond:1-2}}}{=\joinrel=}
        &\;
        \mathscr{O}_{\rm f} \Phi_{\rm p} \mathscr{O}_{\rm p}{}^\dagger (y_{[t-L,t)} - \mathscr{I}_{\rm p} u_{[t-L,t)})
    \end{align*}
    is unique, which implies uniqueness of $y^*_{[t,t+N)}$ via \eqref{Eq:PROOF:LagIniCond:2-1}.
    
    (iii): It follows from \eqref{Eq:LEMMA:LagIniCond:Beh} and \eqref{Eq:LEMMA:LagIniCond:Span} that
    \begin{align*}&
        [u_{[t-L,t)}; u^*_{[t,t+N)}; y_{[t-L,t)}; y^*_{[t,t+N)}]
        \\& \in \mathrm{ColSpan} [U_{\rm p}; U_{\rm f}; Y_{\rm p}; Y_{\rm f}].
    \end{align*}
    By definition of column span, there exists $g \in \real^h$ such that
    \begin{align} \label{Eq:PROOF:LagIniCond:3-0}
        \small \begin{bmatrix}
            U_{\rm p} \\ U_{\rm f} \\ Y_{\rm p} \\ Y_{\rm f}
        \end{bmatrix} g 
        = \begin{bmatrix}
            u_{[t-L,t)} \\ u^*_{[t,t+N)} \\ y_{[t-L,t)} \\ y^*_{[t,t+N)}
        \end{bmatrix}.
    \end{align}
    Now, we show that $Y_{\rm f}$ can be written as $\mathcal{N} [U_{\rm p}; U_{\rm f}; Y_{\rm p}]$ with some matrix $\mathcal{N} \in \real^{pN \times (mL + mN + pL)}$.
    From Lemma \ref{LEMMA:BehColSpan},
    \begin{align*}
        \mathscr{B}^\Sys_{[t-L,t+N)} 
        \;\;\;=\;\;\; &\; \mathrm{ColSpan} \small \left[ \begin{array}{c|c}
            0 & I \\ \hline
            \mathscr{O}_{\rm pf} & \mathscr{I}_{\rm pf}
        \end{array} \right]
        \\
        \overset{\text{via \eqref{Eq:PROOF:LagIniCond:0}}}{=\joinrel=}&\; \small
        \mathrm{ColSpan} \left[\begin{array}{c|cc}
            0 & I & 0 \\
            0 & 0 & I \\ \hline
            \mathscr{O}_{\rm p} & \mathscr{I}_{\rm p} & 0 \\
            \mathscr{O}_{\rm f} \Phi_{\rm p} & \mathscr{O}_{\rm f} \mathscr{C}_{\rm p} & \mathscr{I}_{\rm f}
        \end{array}\right]
    \end{align*}
    Notice that the above column span and the column span in \eqref{Eq:LEMMA:LagIniCond:Span} are both equal to $\mathscr{B}^\Sys_{[t-L,t+N)}$.
    Hence, there exist matrices $\mathcal{P} \in \real^{n\times h}, \mathcal{Q} \in \real^{mL\times h}, \mathcal{R} \in \real^{mN\times h}$ such that
    \begin{align*} \small
        \begin{bmatrix}
            U_{\rm p} \\ U_{\rm f} \\ Y_{\rm p} \\ Y_{\rm f}
        \end{bmatrix}
        = \begin{bmatrix}
            0 & I & 0 \\
            0 & 0 & I \\
            \mathscr{O}_{\rm p} & \mathscr{I}_{\rm p} & 0 \\
            \mathscr{O}_{\rm f} \Phi_{\rm p} & \mathscr{O}_{\rm f} \mathscr{C}_{\rm p} & \mathscr{I}_{\rm f}
        \end{bmatrix}
        \begin{bmatrix}
            \mathcal{P} \\ \mathcal{Q} \\ \mathcal{R}
        \end{bmatrix}.
    \end{align*}
    Computing the right-hand side above and then comparing the result to the left-hand side, we have $\mathcal{Q} = U_{\rm p}$, $\mathcal{R} = U_{\rm f}$, and
    \begin{align}
        \label{Eq:PROOF:LagIniCond:3-1}
        Y_{\rm p} =&\; \mathscr{O}_{\rm p} \mathcal{P} + \mathscr{I}_{\rm p} U_{\rm p}, 
        \\
        \label{Eq:PROOF:LagIniCond:3-2}
        Y_{\rm f} =&\; \mathscr{O}_{\rm f} \Phi_{\rm p} \mathcal{P} + \mathscr{O}_{\rm f} \mathscr{C}_{\rm p} U_{\rm p} + \mathscr{I}_{\rm f} U_{\rm f}.
    \end{align}
    Therefore, we can represent $Y_{\rm f}$ into the form $\mathcal{N} [U_{\rm p}; U_{\rm f}; Y_{\rm p}]$,
    \begin{align} 
    \notag
        Y_{\rm f} 
        \overset{\text{via \eqref{Eq:PROOF:LagIniCond:3-2}}}{=\joinrel=}
        &\; \mathscr{O}_{\rm f} \Phi_{\rm p} \mathcal{P} + \mathscr{O}_{\rm f} \mathscr{C}_{\rm p} U_{\rm p} + \mathscr{I}_{\rm f} U_{\rm f}
    \\ \notag
        \overset{\text{via \eqref{Eq:PROOF:LagIniCond:2-2}}}{=\joinrel=}
        &\; \mathcal{M} \mathscr{O}_{\rm p} \mathcal{P} + \mathscr{O}_{\rm f} \mathscr{C}_{\rm p} U_{\rm p} + \mathscr{I}_{\rm f} U_{\rm f}
    \\ \notag
        \overset{\text{via \eqref{Eq:PROOF:LagIniCond:3-1}}}{=\joinrel=}
        &\;
        \mathcal{M} (Y_{\rm p} - \mathscr{I}_{\rm p} U_{\rm p}) 
        + \mathscr{O}_{\rm f} \mathscr{C}_{\rm p} U_{\rm p} 
        + \mathscr{I}_{\rm f} U_{\rm f}
    \\ \label{Eq:PROOF:LagIniCond:3-3}
        =\;\;\; &\;
        \mathcal{N} [U_{\rm p}; U_{\rm f}; Y_{\rm p}]
    \end{align}
    with matrix $\mathcal{N} := [\mathscr{O}_{\rm f} \mathscr{C}_{\rm p}-\mathcal{M}\mathscr{I}_{\rm p}, \mathscr{I}_{\rm f}, \mathcal{M}]$.
    The result \eqref{Eq:LEMMA:LagIniCond:Repre} then follows:
    \begin{align*}
        y^*_{[t,t+N)} 
        \overset{\text{via \eqref{Eq:PROOF:LagIniCond:3-0}}}{=\joinrel=}
        &\; Y_{\rm f} g
        \overset{\text{via \eqref{Eq:PROOF:LagIniCond:3-3}}}{=\joinrel=}
        \mathcal{N} \mathcal{H} g
        = \mathcal{N} \mathcal{H} \mathcal{H}^\dagger \mathcal{H} g
        \\
        \overset{\text{via \eqref{Eq:PROOF:LagIniCond:3-3}}}{=\joinrel=} 
        &\; Y_{\rm f} \mathcal{H}^\dagger \mathcal{H} g
        \overset{\text{via \eqref{Eq:PROOF:LagIniCond:3-0}}}{=\joinrel=}
        Y_{\rm f} \mathcal{H}^\dagger \small \begin{bmatrix}
            u_{[t-L,t)} \\ u^*_{[t,t+N)} \\ y_{[t-L,t)}
        \end{bmatrix},
    \end{align*}
    where we let $\mathcal{H}$ denote $[U_{\rm p}; U_{\rm f}; Y_{\rm p}]$.
    \end{subequations}
\end{pfof}

\begin{pfof}{Lemma \ref{LEMMA:LiftEqv}}
    Let \eqref{Eq:LTV} be the model of $\Sys$, and \eqref{Eq:LiftSys} be the model of $\LiftSys(t_0)$.
    Define matrices $\mathscr{O}^{t_2}_{t_1}$ and $\mathscr{I}^{t_2}_{t_1}$ (resp. $\mathscr{\hat O}^{t_2}_{t_1}$ and $\mathscr{\hat I}^{t_2}_{t_1}$) via \eqref{Eq:SysMatNotation_O} and \eqref{Eq:SysMatNotation_I} for system $\Sys$ (resp. $\LiftSys(t_0)$).
    By Lemma \ref{LEMMA:BehColSpan}, the respective behaviors can be expressed as
    \begin{equation}\label{Eq:PROOF:LiftEqv}
    \begin{aligned}
        \mathscr{B}^\Sys_{[t_0,\, t_0 + s T)} =&\; \mathrm{ColSpan} \small \begin{bmatrix}
            0 & I \\ \mathscr{O}^{t_0+sT-1}_{t_0} 
            & \mathscr{I}^{t_0+sT-1}_{t_0}
        \end{bmatrix},
        \\
        \mathscr{B}^{\LiftSys(t_0)}_{[0, s)} =&\; \mathrm{ColSpan} \small \begin{bmatrix}
            0 & I \\ 
            \mathscr{\hat O}^{s-1}_0 
            & \mathscr{\hat I}^{s-1}_0
        \end{bmatrix}.
    \end{aligned}
    \end{equation}
    Note that $\mathscr{\hat O}^{s-1}_0$ and $\mathscr{\hat I}^{s-1}_0$ are defined based on $\liftmat{A}, \liftmat{B}, \liftmat{C}, \liftmat{D}$.
    Using the definitions of $\liftmat{A}, \liftmat{B}, \liftmat{C}, \liftmat{D}$ from \eqref{Eq:LiftedMatrices} and the periodicity of $\Sys$, one can verify that for all $s \in \natural$
    \begin{align*}
        \mathscr{O}^{t_0+sT-1}_{t_0} = \mathscr{\hat O}^{s-1}_0, \qquad
        \mathscr{I}^{t_0+sT-1}_{t_0} = \mathscr{\hat I}^{s-1}_0,
    \end{align*}
    and hence the column spans in \eqref{Eq:PROOF:LiftEqv} are equal. 
\end{pfof}

\begin{pfof}{Lemma \ref{LEMMA:OrderLagLift_LTP}}
    \begin{subequations}
    Let \eqref{Eq:LTV} be the model of $\Sys$, and \eqref{Eq:LiftSys} be the model of $\LiftSys(t)$.
    Define matrix $\mathscr{O}^{t_2}_{t_1}$ (resp. $\mathscr{\hat O}^{t_2}_{t_1}$) via \eqref{Eq:SysMatNotation_O} for system $\Sys$ (resp. $\LiftSys(t)$).
    Using the definitions of $\liftmat{A}$ and $\liftmat{C}$ from \eqref{Eq:LiftedMatrices} and the periodicity of $\Sys$, one can verify that
    \begin{align} \label{Eq:PROOF:OrderLagLift_LTP:1}
        \mathscr{\hat O}^{s-1}_0
        = \mathscr{O}^{t+sT-1}_t
        \qquad \forall s \in \natural.
    \end{align}
    Let $a_s := \rank(\mathscr{O}^{t+s-1}_t)$ and $b_s := \rank(\mathscr{\hat O}^{s-1}_0)$.
    It follows from \eqref{Eq:PROOF:OrderLagLift_LTP:1} that $b_s = a_{sT}$.
    By \eqref{Eq:OrderLag_LTV}, we now compute that
    \begin{align*}
        \order(\LiftSys(t)) 
        = \lim_{s \to \infty} b_s
        = \lim_{s \to \infty} a_{sT}
        = \lim_{s' \to \infty} a_{s'}
        = \order(\Sys, t),
    \end{align*}
    which shows (i).
    Similarly, we compute from \eqref{Eq:OrderLag_LTV} that
    \begin{align}
    \begin{split} \label{Eq:PROOF:OrderLagLift_LTP:2}
        \lag(\LiftSys(t))
        =&\; \min \{ s\in\natural : b_s = \order(\LiftSys(t)) \}
        \\=&\; \min \{ s\in\natural : a_{sT} = \order(\Sys, t) \},
    \end{split}
        \\ \label{Eq:PROOF:OrderLagLift_LTP:3}
        \lag(\Sys, t)
        =&\; \min \{ s'\in\natural : a_{s'} = \order(\Sys, t) \}.
    \end{align}
    We know from the minimality in \eqref{Eq:PROOF:OrderLagLift_LTP:2} that
    \begin{align*}
        a_{[\lag(\LiftSys(t))-1]T} \neq \order(\Sys, t), \qquad
        a_{\lag(\LiftSys(t))T} = \order(\Sys, t).
    \end{align*}
    By the minimality in \eqref{Eq:PROOF:OrderLagLift_LTP:3}, the above relations imply that
    \begin{align*}
        [\lag(\LiftSys(t))-1]T < \lag(\Sys, t) \leq \lag(\LiftSys(t))T,
    \end{align*}
    which is equivalent to (ii).
    \end{subequations}
\end{pfof}

\begin{pfof}{Corollary \ref{COROLLARY:OrderLagBound_LTP}}
    (i) follows from \eqref{Eq:OrderLag_LTV}, as $\rank(\mathscr{O}^{t_2}_{t_1})$ is bounded by the row size $n$ of $\mathscr{O}^{t_2}_{t_1}$.
    (ii) is shown below:
    \begin{align*}
        \lag(\Sys,t) 
        \leq \lag(\LiftSys(t)) T 
        \leq \order(\LiftSys(t)) T 
        = \order(\Sys,t) T 
        \leq n T,
    \end{align*}
    where the first inequality is by Lemma \ref{LEMMA:OrderLagLift_LTP}(ii), and the second inequality is because $\lag(\LiftSys(t)) \leq \order(\LiftSys(t))$ for the LTI system $\LiftSys(t)$ \cite[Eq. (2)]{BEHAV:Markovsky2021}.
\end{pfof}

\begin{pfof}{Lemma \ref{LEMMA:CtrbEqv}}
    Let $T$ be the period of $\Sys$, and let $\liftvec{w}_{[\tau_1,\tau_2]} := [\liftvec{u}_{[\tau_1,\tau_2]}; \liftvec{y}_{[\tau_1,\tau_2]}]$ denote a trajectory of the lifted system $\LiftSys(t_0)$.
    Given a trajectory $w_{[t_0,\infty)}$ of $\Sys$, let $\liftvec{w}_{[0,\infty)}$ be the trajectory of $\LiftSys(t_0)$ equal to $w_{[t_0,\infty)}$, and vise versa,
    where we implicitly used $\mathscr{B}^\Sys_{[t_0,\infty)} = \mathscr{B}^{\LiftSys(t_0)}_{[0,\infty)}$ via Lemma \ref{LEMMA:LiftEqv}.
    
    \textbf{If.} For $t_0 \in \integer$, consider trajectories $w^\text{I}_{[t_0,\infty)}, w^\text{II}_{[t_0,\infty)} \in \mathscr{B}^\Sys_{[t_0,\infty)}$ and an arbitrary integer $t_1 \geq t_0$.
    Choose some $\tau_1 \in \integer$ such that $t_0 + \tau_1 T \geq t_1$.
    Since $\liftvec{w}^\text{I}_{[0,\infty)}$ and $\liftvec{w}^\text{II}_{[0,\infty)}$ are trajectories of $\LiftSys(t_0)$, by controllability of $\LiftSys(t_0)$,
    there exists an integer $\tau_2 \geq \tau_1$ and a trajectory $\liftvec{w}^\diamond_{[0,\infty)} \in \mathscr{B}^{\LiftSys(t_0)}_{[0,\infty)}$ such that
    \begin{align} \label{Eq:PROOF:CtrbLiftEqv}
        \liftvec{w}^\diamond_{[0,\tau_1)}
        = \liftvec{w}^\text{I}_{[0,\tau_1)}, \qquad
        \liftvec{w}^\diamond_{[\tau_2,\infty)}
        = \liftvec{w}^\text{II}_{[\tau_2,\infty)}
    \end{align}
    Let $t_2 := t_0 + \tau_2 T$.
    From \eqref{Eq:PROOF:CtrbLiftEqv} and the equivalence of $w$ and $\liftvec{w}$, $w^\diamond_{[t_0,\infty)}$ is such a trajectory that \eqref{Eq:DEF:Ctrb} holds.
    Since $t_0$ and $t_1$ are arbitrary, $\Sys$ is controllable by Definition \ref{DEF:Ctrb}.
    
    \textbf{Only if.} Consider trajectories $\liftvec{w}^\text{I}_{[0,\infty)}, \liftvec{w}^\text{II}_{[0,\infty)} \in \mathscr{B}^{\LiftSys(t_0)}_{[0,\infty)}$ and an arbitrary integer $\tau_1 \geq 0$.
    Since $w^\text{I}_{[t_0,\infty)}$ and $w^\text{II}_{[t_0,\infty)}$ are trajectories of $\Sys$, from controllability of $\Sys$ there exists an integer $t_2 \geq t_1 := t_0 + \tau_1 T$ and a trajectory $w^\diamond_{[t_0,\infty)} \in \mathscr{B}^\Sys_{[t_0, \infty)}$ such that \eqref{Eq:DEF:Ctrb} holds.
    Choose some $\tau_2 \in \integer$ satisfying $t_0 + \tau_2 T \geq t_2$.
    From \eqref{Eq:DEF:Ctrb} and the equivalence of $w$ and $\liftvec{w}$, the trajectory $\liftvec{w}^\diamond_{[0,\infty)}$ satisfies \eqref{Eq:PROOF:CtrbLiftEqv}.
    Since $\tau_1$ is arbitrary, the LTI system $\LiftSys(t_0)$ is controllable by Definition \ref{DEF:Ctrb}.
\end{pfof}

\begin{pfof}{Lemma \ref{LEMMA:Fund_LTP}}
    \begin{subequations}
    We first prove the case when $K$ is a multiple of $T$, i.e., where $K = K_1 T$ for some $K_1 \in \natural$. Let 
    \begin{align*}
        \liftvec{w}_{[\tau_1, \tau_2]} := [\liftvec{u}_{[\tau_1, \tau_2]}; \liftvec{y}_{[\tau_1, \tau_2]}]
    \end{align*}
    denote a trajectory of the lifted system $\LiftSys(t_1)$.
    From Lemma \ref{LEMMA:LiftEqv}, we know that $\mathscr{B}^{\LiftSys(t_1)}_{[0,s)} = \mathscr{B}^\Sys_{[t_1, t_1 + sT)}$ for all $s \in \natural$, so we can establish such a trajectory $\liftvec{w}^{\rm d}_{[0,P)} \in \mathscr{B}^{\LiftSys(t_1)}_{[0,P)}$ that
    \begin{align*}
        \liftvec{w}^{\rm d}_{[0,P)} := w^{\rm d}_{[t_1, t_1+PT)}
    \end{align*}
    with $P := \lfloor (t_2-t_1+1)/T \rfloor$, i.e., $P$ is the number of whole periods in the interval $[t_1, t_2]$.
    With abuse of notation, we let $\order := \order(\Sys, t_1)$ and $\order_\mathsf{L} := \order(\LiftSys(t_1))$ in this proof, then we have $\order = \order_\mathsf{L}$ via Lemma \ref{LEMMA:OrderLagLift_LTP}(i).
    By direct substitution, one can verify that\footnote{Note that in \eqref{Eq:PROOF:Fund_LTP:1}, the block rows on the left-hand side are of size $mT$, where $m$ is the input dimension of $\Sys$, while on the right-hand side, the block rows are of size $m$. Similarly for \eqref{Eq:PROOF:Fund_LTP:2}.}
    \begin{align} 
        \label{Eq:PROOF:Fund_LTP:1}
        \mathcal{H}_{K_1+\order_\mathsf{L}}(\liftvec{u}^{\rm d}_{[0,P)})
        =&\; \mathcal{H}^T_{(K_1+\order)T}(u^{\rm d}_{[t_1, t_2]}),
        \\
        \label{Eq:PROOF:Fund_LTP:2}
        \mathcal{H}_{K_1} (\liftvec{w}^{\rm d}_{[0,P)})
        =&\; \mathcal{H}^T_K (w^{\rm d}_{[t_1,t_2]}).
    \end{align}
    Since $u^{\rm d}_{[t_1,t_2]}$ is $T$-p.p.e. of order $(K_1 + \order) T$ (i.e., the right-hand side of \eqref{Eq:PROOF:Fund_LTP:1} has full row rank),
    we know that $\liftvec{u}^{\rm d}_{[0,P)}$ is p.e. of order $K_1 + \order_\mathsf{L}$ (as the left-hand side of \eqref{Eq:PROOF:Fund_LTP:1} has full row rank).
    We also know via Lemma \ref{LEMMA:CtrbEqv} that $\LiftSys(t_1)$ is controllable because $\Sys$ is controllable.
    Thus by Lemma \ref{LEMMA:Fund} we have
    \begin{align} \label{Eq:PROOF:Fund_LTP:3}
        \footnotesize
        \mathrm{ColSpan} (\mathcal{H}_{K_1} (\liftvec{w}^{\rm d}_{[0,P)}))
        = \mathscr{B}^{\LiftSys(t_1)}_{[0, K_1)}.
    \end{align}
    Substitute \eqref{Eq:PROOF:Fund_LTP:2} into the left-hand side of \eqref{Eq:PROOF:Fund_LTP:3}, and replace the right-hand side of \eqref{Eq:PROOF:Fund_LTP:3} using $\mathscr{B}^{\LiftSys(t_1)}_{[0, K_1)} = \mathscr{B}^{\Sys}_{[t_1, t_1 + K)}$ (via Lemma \ref{LEMMA:LiftEqv}),
    and then we obtain the result \eqref{Eq:LEMMA:Fund_LTP}.
    
    Next, we show the result for all $K \in \natural$.
    Let $K_1 := \lceil K/T \rceil$ and $\widehat K := K_1 T$, i.e., $\widehat K$ is the smallest multiple of $T$ greater than or equal to $K$.
    Since $\lceil K/T\rceil = \lceil \widehat{K}/T\rceil$, $u^{\rm d}_{[t_1,t_2]}$ is $T$-p.p.e. of order $(\lceil \widehat{K}/T\rceil + \order(\Sys,t_1)) T$, we have the condition (ii) of this lemma for the case $K \gets \widehat K$ (as a multiple of $T$), which case we have already proved. We therefore have
    \begin{align} \label{Eq:PROOF:Fund_LTP:4}
        \mathrm{ColSpan} (\mathcal{H}^T_{\widehat K} (w^{\rm d}_{[t_1, t_2]}))
        = \mathscr{B}^\Sys_{[t_1, t_1 + \widehat K)}.
    \end{align}
    Define $\mathcal{H}_\triangle := \mathcal{H}^T_K(w^{\rm d}_{[t_1,t_2-(\widehat K - K)]})$.
    One can verify that $\mathcal{H}^T_K (u^{\rm d}_{[t_1,t_2-(\widehat K - K)]})$ consists of the first $K$ block rows of $\mathcal{H}^T_{\widehat K} (u^{\rm d}_{[t_1,t_2]})$, and similarly for $y^{\rm d}$.
    Hence, applying Lemma \ref{LEMMA:SubBehRepre} for \eqref{Eq:PROOF:Fund_LTP:4} we have
    \begin{equation}\label{Eq:PROOF:Fund_LTP:5}
        \mathrm{ColSpan}(\mathcal{H}_\triangle)
        = \mathscr{B}^\Sys_{[t_1, t_1 + K)}.
    \end{equation}
    However, note that
    \begin{equation}\label{Eq:PROOF:Fund_LTP:6}
        \mathrm{ColSpan}(\mathcal{H}_\triangle)
        \subseteq \mathrm{ColSpan}(\mathcal{H}^T_K (w^{\rm d}_{[t_1,t_2]}))
        \subseteq \mathscr{B}^\Sys_{[t_1, t_1 + K)},
    \end{equation}
    where the first inclusion ($\subseteq$) above is because $\mathcal{H}_\triangle$ is a sub-matrix of $\mathcal{H}^T_K (w^{\rm d}_{[t_1,t_2]})$ with the same column size, and the second inclusion above is because each column of $\mathcal{H}^T_K (w^{\rm d}_{[t_1,t_2]})$ is a vector in the behavior set $\mathscr{B}^\Sys_{[t_1, t_1 + K)}$. 
    The equality \eqref{Eq:LEMMA:Fund_LTP} now follows by combining \eqref{Eq:PROOF:Fund_LTP:5} and \eqref{Eq:PROOF:Fund_LTP:6}.
    \end{subequations}
\end{pfof}

\begin{pfof}{Lemma \ref{LEMMA:DataSpan}}
    Let $t_{\rm d1}^\Idx := t_{\rm d1}+(\Idx-1)$ and $t_{\rm d2}^\Idx := t_{\rm d2}-(T-\Idx)$, and notice that
    \begin{align*}
        [U_{\rm p}^\Idx; U_{\rm f}^\Idx; Y_{\rm p}^\Idx; Y_{\rm f}^\Idx] 
        = \mathcal{H}^T_K(w^{\rm d}_{[t_{\rm d1}^\Idx, t_{\rm d2}^\Idx]})
    \end{align*}
    Also let $H := (\lceil (L+N)/T \rceil + \order(\Sys, t_{\rm d1})) T$ and $\widehat H := (\lceil K/T \rceil + \order(\Sys, t_{\rm d1})) T$.
    
    We first conclude that $u^{\rm d}_{[t_{\rm d1}^\Idx, t_{\rm d2}^\Idx]}$ is $T$-p.p.e of order $H$, i.e., $\mathcal{H}^T_H(u^{\rm d}_{[t_{\rm d1}^\Idx, t_{\rm d2}^\Idx]})$ has full row rank. This is because $\mathcal{H}^T_{\widehat H}(u^{\rm d}_{[t_{\rm d1}, t_{\rm d2}]})$ has full row rank (since $u^{\rm d}_{[t_{\rm d1}, t_{\rm d2}]}$ is $T$-p.p.e of order $\widehat H$) and $\mathcal{H}^T_H(u^{\rm d}_{[t_{\rm d1}^\Idx, t_{\rm d2}^\Idx]})$ is a sub-matrix of $\mathcal{H}^T_{\widehat H}(u^{\rm d}_{[t_{\rm d1}, t_{\rm d2}]})$.
    
    Then, the result \eqref{Eq:LEMMA:DataSpan} follows from Lemma \ref{LEMMA:Fund_LTP}, with controllability of $\Sys$ and periodic persistent excitation of $u^{\rm d}_{[t_{\rm d1}^\Idx, t_{\rm d2}^\Idx]}$.
\end{pfof}

\begin{pfof}{Proposition \ref{PROP:Perfm}}
    \textbf{Equivalence of Optimal Sets.}
    We first show that the problems \eqref{PROB:DeePC_Prob}, \eqref{PROB:SPC_Prob} and \eqref{PROB:MPC_Prob} have the same set of optimal trajectories $w^*_{[t,t+N)}$.
    Since the three problems have the same cost function \eqref{Eq:Cost} and a common constraint \eqref{Eq:I/O_Cons},
    it suffices to show the rest constraints, i.e. the following statements, are equivalent:
    \begin{enumerate}[a)]
        \item $u^*_{[t,t+N)}$ and $y^*_{[t,t+N)}$ satisfy \eqref{Eq:DeePC_Cons} for some $g \in \real^h$;
        \item $u^*_{[t,t+N)}$ and $y^*_{[t,t+N)}$ satisfy \eqref{Eq:SPC_Cons};
        \item $u^*_{[t,t+N)}$ and $y^*_{[t,t+N)}$ satisfy \eqref{Eq:MPC_Cons} for some $x^*_{[t,t+N]}$.
    \end{enumerate}
    With (ii) and (iii), we obtain \eqref{Eq:OfflineData_OnlineBeh} as discussed in Section \ref{SUBSEC:Algo:OfflineData}, where we used Lemma \ref{LEMMA:DataSpan} and the definition of the proper index $\ProperIdx(t)$. It follows from (iv) and \eqref{Eq:OfflineData_OnlineBeh} that
    \begin{align} \label{Eq:PROOF:Perfm:DataMatSpan}
        \mathrm{ColSpan} [U_{\rm p}^{\EstProperIdx(t)}; U_{\rm f}^{\EstProperIdx(t)}; Y_{\rm p}^{\EstProperIdx(t)}; Y_{\rm f}^{\EstProperIdx(t)}]
        = \mathscr{B}^\Sys_{[t-L, t+N)}.
    \end{align}
    For showing the equivalence of a), b) and c), we introduce an auxiliary statement:
    \begin{enumerate}[a)] \setcounter{enumi}{3}
        \item $u^*_{[t,t+N)}$ and $y^*_{[t,t+N)}$ satisfy \eqref{Eq:LEMMA:LagIniCond:Beh}.
    \end{enumerate}
    
    a) $\iff$ d):
    By definition of column span, a) is same as
    \begin{align*}
        \begin{bmatrix}
            u_{[t-L,t)} \\ u^*_{[t,t+N)} \\ y_{[t-L,t)} \\ y^*_{[t,t+N)}
        \end{bmatrix}
        \in \mathrm{ColSpan} \footnotesize
        \begin{bmatrix}
            U_{\rm p}^{\EstProperIdx(t)} \\ U_{\rm f}^{\EstProperIdx(t)} \\ Y_{\rm p}^{\EstProperIdx(t)} \\ Y_{\rm f}^{\EstProperIdx(t)}
        \end{bmatrix}.
    \end{align*}
    From \eqref{Eq:PROOF:Perfm:DataMatSpan}, the expression above is equivalent to d). 
    
    b) $\iff$ d):
    Given (i) and \eqref{Eq:PROOF:Perfm:DataMatSpan}, via Lemma \ref{LEMMA:LagIniCond}(iii), 
    for each $u^*_{[t,t+N)}$ \eqref{Eq:SPC_Cons} specifies the unique $y^*_{[t,t+N)}$ that satisfies \eqref{Eq:LEMMA:LagIniCond:Beh}.
    Therefore, b) is equivalent to d).
    
    c) $\implies$ d):
    Since $w^*_{[t,t+N)}$ is a trajectory with initial state $x_t$ via \eqref{Eq:MPC_Cons} and $w_{[t-L,t)}$ is a trajectory with final state $x_t$, the two trajectories can be connected into a single trajectory, i.e. \eqref{Eq:LEMMA:LagIniCond:Beh} holds.
    
    d) $\implies$ c):
    Define $y^{**}_{[t,t+N)}$ the unique output resulting from the initial state $x_t$ and input $u^*_{[t,t+N)}$.
    Since $w_{[t-L,t)}$ is a trajectory with final state $x_t$ and $[u^*_{[t,t+N)}; y^{**}_{[t,t+N)}]$ is a trajectory with initial state $x_t$, their connection is also a trajectory and satisfies the following.
    \begin{align*}
        [u_{[t-L,t)}; u^*_{[t,t+N)}; y_{[t-L,t)}; y^{**}_{[t,t+N)}] \in \mathscr{B}^\Sys_{[t-L,t+N)}
    \end{align*}
    Comparing \eqref{Eq:LEMMA:LagIniCond:Beh} to the above, due to the uniqueness in Lemma \ref{LEMMA:LagIniCond}(ii) (where we used (i) of this proposition), we conclude that $y^*_{[t,t+N)} = y^{**}_{[t,t+N)}$.
    Hence, $y^*_{[t,t+N)}$ is the output resulting from the initial state $x_t$ and input $u^*_{[t,t+N)}$, i.e., \eqref{Eq:MPC_Cons} holds.
    
    \textbf{Uniqueness of Future Trajectory.}
    Finally, we show that the optimal trajectory $w^*_{[t,t+N)}$ of each problem is unique.
    \eqref{PROB:MPC_Prob} has a unique optimal solution $w^*_{[t,t+N)}$, because $x^*_{[t,t+N]}$ and $y^*_{[t,t+N)}$ are both dependent on $u^*_{[t,t+N)}$ via \eqref{Eq:MPC_Cons} and hence the cost function \eqref{Eq:Cost} with $R \succ 0$ is strictly convex of the only independent variable $u^*_{[t,t+N)}$.
    Following from the equivalence of the optimal sets, the optimal trajectories $w^*_{[t,t+N)}$ of \eqref{PROB:DeePC_Prob} and \eqref{PROB:SPC_Prob} are also unique.
    This completes the proof.
\end{pfof}

\end{document}